\documentclass[pdflatex,sn-mathphys-num]{sn-jnl}

\usepackage{graphicx}%
\usepackage{multirow}%
\usepackage{amsmath,amssymb,amsfonts}%
\usepackage{amsthm}%
\usepackage{mathrsfs}%
\usepackage[title]{appendix}%
\usepackage{xcolor}%
\usepackage{textcomp}%
\usepackage{manyfoot}%
\usepackage{booktabs}%
\usepackage{algorithm}%
\usepackage{algorithmicx}%
\usepackage{algpseudocode}%
\usepackage{listings}%
\usepackage{upgreek}

\theoremstyle{thmstyleone}%
\theoremstyle{thmstyletwo}%

\theoremstyle{thmstylethree}%

\begin{document}

\title[]{A wafer-scale ultrasensitive programmable chiroptical sensor}

\author[1]{\fnm{Haoyu} \sur{Xie}}
\author[1]{\fnm{Jichao} \sur{Fan}}
\author[2]{\fnm{Zarif Ahmad Razin} \sur{Bhuiyan}}
\author[2]{\fnm{Saqlain} \sur{Raza}}
\author[3]{\fnm{Mohammad} \sur{Mohammadi}}
\author[4]{\fnm{Cheng} \sur{Guo}}
\author[3]{\fnm{Yunshan} \sur{Wang}}
\author[2]{\fnm{Jun} \sur{Liu}}
\author*[1]{\fnm{Weilu} \sur{Gao}}\email{weilu.gao@utah.edu}

\affil[1]{Department of Electrical and Computer Engineering, The University of Utah, Salt Lake City, UT, USA}

\affil[2]{Department of Mechanical and Aerospace Engineering, North Carolina State University, Raleigh, NC, USA}

\affil[3]{Department of Chemical Engineering, The University of Utah, Salt Lake City, UT, USA}

\affil[4]{Department of Electrical and Computer Engineering, The University of Texas at Austin, Austin, TX, USA}

\abstract{Chiroptical enantioselective sensing is gaining traction across various applications. However, intrinsic molecular chiroptical responses are weak, and existing amplification approaches add synthesis, manufacturing, or operational complexity that limits sensitivity, scalability, and dynamic control. Here, we present a fundamentally new sensing paradigm merging adsorption-driven chirality induction with wafer-scale optical transduction in a programmable heterostructure containing twisted aligned carbon nanotubes (CNTs) and phase change materials (PCMs). Chiral molecules adsorb onto CNTs to form chiroptically active composites that are macroscopically assembled by alignment and rotational stacking, yielding large ultraviolet circular dichroism (CD). We resolve molecule concentration and handedness in a single device without lithography, hotspot delivery, or differential protocols, achieving sub-\unboldmath$\upmu$M sensitivity for CD-silent glucose and chiral amino acids enabled by \unboldmath$>10^5\,\mathrm{M^{-1}}$ adsorption constants. We validate adsorption using molecular dynamics simulations, reproduce experimental results using chiral transfer matrix simulations, and realize sensor programmability by tuning the PCM layer. This platform enables cost-effective in-situ enantiomer monitoring in aqueous environments.}

\maketitle

\section*{Introduction}

Chiral molecules, existing as two non-superimposable mirror-image forms called enantiomers, are ubiquitous. Their handedness, either right (D, \textit{dexter}) or left (L, \textit{laevus}), governs critical functions across a variety of natural and technological contexts~\cite{BarronEtAl2009}. Accordingly, enantioselective sensing, the ability to distinguish and quantify two enantiomers, particularly at trace levels, is essential for ensuring correct molecular function and informed decision-making in applications where even minute enantiomeric imbalances are decisive. For example, in pharmaceuticals, two enantiomers of a drug can exhibit sharply different biological effects, making stringent control of enantiomeric excess and trace impurities vital to safety and efficacy~\cite{NguyenEtAl2006IJBSI,IzakeEtAl2007JPS}. In clinical diagnostics and personalized medicine, abnormal enantiomeric ratios of chiral small molecules in biofluids and tissues are increasingly linked to diseases ranging from cancers to kidney and brain disorders, positioning these molecules as promising biomarkers. However, their low concentrations demand sensitive and reliable assays that integrate seamlessly with real-world clinical workflows~\cite{LiuEtAl2023NRC}. In food systems, enantiomeric signatures provide rapid, information-rich markers of contamination, and the widespread use of chiral pesticides drives the need for enantioselective residue monitoring to accurately assess exposure and risk~\cite{JeschkeEtAl2018PMS,WangEtAl2023JAFC}.

Conventional analytical enantioselective sensing approaches based on chiral agents or chiral separation, such as chromatography, nuclear magnetic resonance, and mass spectrometry, are powerful but resource-intensive, slow, and poorly suited to continuous, in situ monitoring, particularly in aqueous media~\cite{PenasaEtAl2025CSR}. Electrochemical and field-effect-transistor-like chiral sensors offer attractive routes toward miniaturized, low-power devices, yet they face practical barriers. Electrochemical approaches are only effective for electroactive analytes, and when involving antibodies, can be costly and time-consuming~\cite{ZorEtAl2019TTAC}. Electronic approaches often rely on specialized organic semiconductors and require electrical contacts and routing~\cite{TorsiEtAl2008NM}, increasing material and packaging complexity for continuous operation in aqueous and complex environments. In contrast, chiroptical spectroscopy, such as circular dichroism (CD) spectroscopy that captures the differential optical absorption between left- and right-handed circularly polarized light, can provide rapid, cost-effective, non-destructive, multiplexed, and high-throughput detection, while suffering from low sensitivity because of weak or even silent molecular chiroptical responses. Two broad amplification strategies have therefore emerged. Molecule-level induced-CD approaches, including supramolecular and host-guest systems, produce measurable chiroptical signals when chiral molecules bind to and interact with achiral chromophoric reporters~\cite{AllenmarkEtAl2003C,HemburyEtAl2008CR,WolfEtAl2013CSR,FormenEtAl2024ACIE}. However, achieving strong responses demands intricate molecular design and synthesis, which can limit robustness, scalability, and deployability. Device-level optical-hotspot approaches, exemplified by resonant metamaterials and plasmonic nanostructures, create enhanced optical near fields in engineered asymmetric structures~\cite{LuoEtAl2017AOM,YooEtAl2019N,MunEtAl2020LSA,WarningEtAl2021N,ImEtAl2024N}. However, these platforms often require sophisticated nanofabrication. This is particularly challenging in the ultraviolet, where many chiral molecules exhibit strong spectral signatures, because of subwavelength feature sizes and limited suitable materials. They also require efficient delivery of chiral molecules into nanoscale hotspots and often rely on differential protocols that compare left- and right-handed devices to suppress common-mode backgrounds, all of which complicates scalable implementation. Moreover, it remains challenging to achieve dynamically tunable enantioselective sensing performance at a wafer scale with energy-efficient electrical programmability and reliable operation for practical deployment, as most current schemes are based on chemical triggers~\cite{LuoEtAl2017AOM,WarningEtAl2021N}.

Here, we present a new enantioselective sensing paradigm that merges molecule-level chirality induction and device-level optical transduction, and demonstrate a wafer-scale, ultrasensitive, programmable chiroptical sensor built as a layered heterostructure consisting of twisted stacks of adsorption-mediated aligned carbon nanotube (CNT) films and tunable optical materials. At the molecular level, chiral molecules adsorb onto CNTs and transform achiral CNTs into a chiroptically active hybrid composite. At the device level, these nano-objects are self-assembled into macroscopically aligned films that are rotationally stacked and integrated with dielectric spacers and tunable optical layers to form a twisted multilayer architecture, collectively amplifying and transducing the responses of nanoscale composites into large CD signals at a wafer scale. Notably, CNTs exhibit ultrastrong ultraviolet CD responses arising from quantum-confinement-driven optical resonances, providing access to the ultraviolet spectral regime where most chiral molecules have prominent features and suitable engineered structures are scarce. We fabricate wafer-scale aligned CNT films using a simple solution-based self-assembly technique and define twist angles through a controlled, deterministic rotational stacking, avoiding complex synthesis and sophisticated nanofabrication. We focus on CD-silent glucose molecules as a stringent test, while also demonstrating generality to chiral amino acids. We experimentally demonstrate continuous, in situ enantioselective sensing, which simultaneously determines concentration and handedness of enantiomers, using a single device without requiring analyte delivery into hotspots or a differential common-mode suppression protocol. We achieve an ultrasensitive sub-$\upmu$M limit of detection (LOD) thanks to consistently large $>10^5\,\mathrm{M^{-1}}$ adsorption equilibrium constants. Further, we corroborate molecule adsorption on CNTs using molecular dynamics (MD) simulations, and reproduce experimental CD spectra and enantioselective sensing curves through optical simulations using a chiral transfer matrix (cTMM) method, both showing excellent agreement with experiments. In addition, we demonstrate programmable sensing by inserting and tuning a phase change material (PCM) film between twisted aligned CNT layers. The demonstrated scalable, ultrasensitive, and programmable chiroptical sensing platform that converts weak molecular handedness into strong CD responses enables real-time enantiomer monitoring in practical aqueous environments across domains ranging from biomedicine and pharmaceutical manufacturing to food safety and environmental surveillance.

\section*{Results}

\subsection*{Sensor architecture and characterization}

\begin{figure}[hbt]
    \centering
    \includegraphics[width=\textwidth]{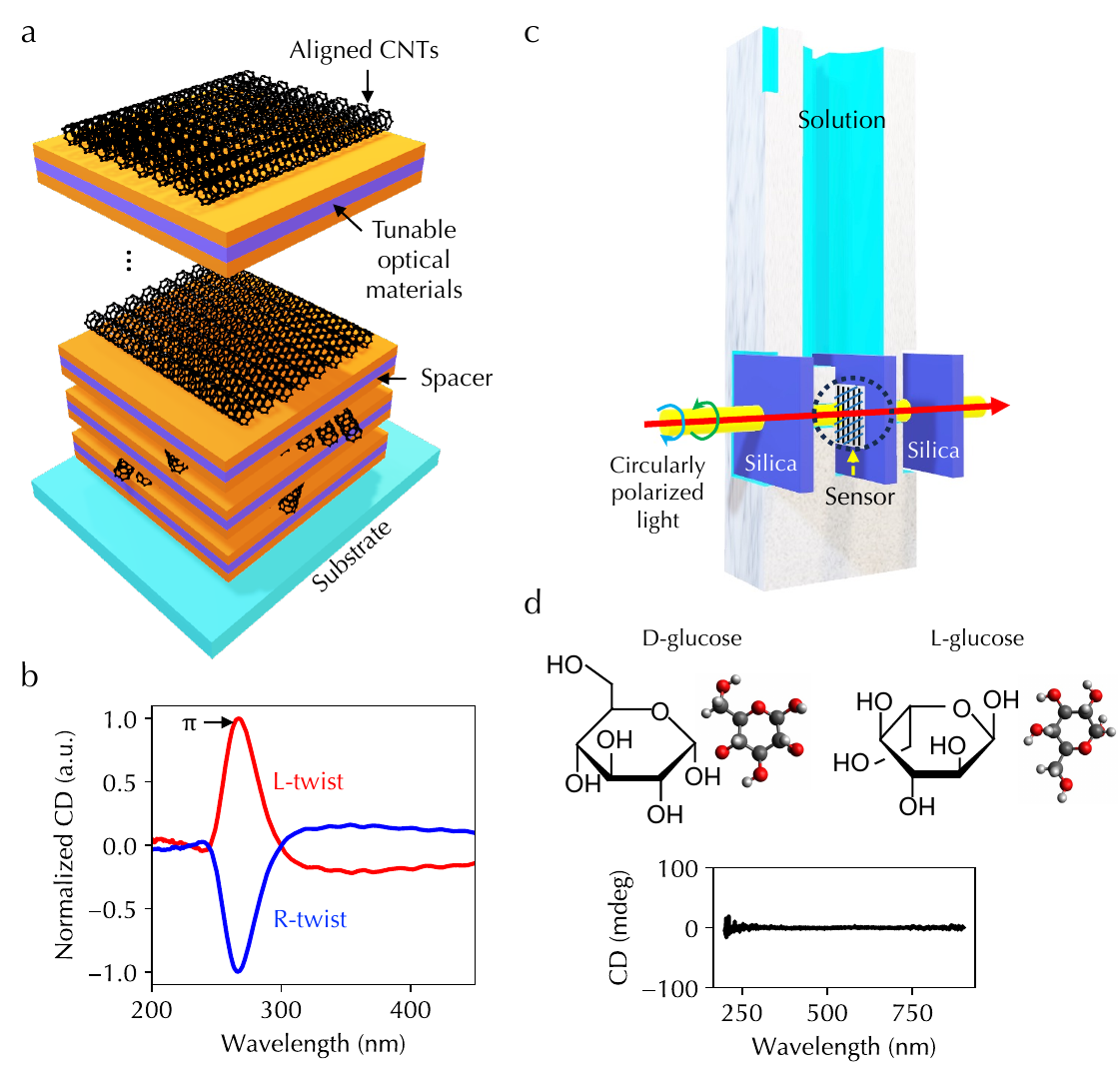}
    \vspace{-15pt}
    \caption{\textbf{Sensor architecture and characterization}. (a)~Schematic diagram of the sensor architecture consisting of multiple layers of twisted aligned CNTs, dielectrics, and tunable optical materials. (b)~Normalized CD spectra of two-layer left-handed (L-twist, red line) and right-handed (R-twist, blue line) twisted aligned CNTs with a $40^{\circ}$ rotation angle between layers. (c)~Schematic illustration of the apparatus for in situ characterizing sensors in aqueous solutions of chiral molecules. (d)~Structural formula of D- and L-glucose molecules in water and its featureless CD spectrum.}
    \label{fig:sensor_arc}
\end{figure}

Figure\,\ref{fig:sensor_arc}a schematically illustrates the chiroptical sensor architecture, which consists of multiple layers of macroscopically aligned CNT films with different in-plane alignment directions, dielectric spacers, and tunable optical materials. The CNTs in the films are achiral, containing a racemic mixture without any chiroptical responses at the individual CNT level. Instead, because of the helical arrangement of one-dimensional (1D) dipole moments in aligned CNTs, twisted aligned CNT structures exhibit designable structure-induced chirality and chiroptical responses, which can be further dynamically programmed by incorporating tunable optical materials~\cite{DoumaniEtAl2023NC,FanEtAl2025NC}. Specifically, we employed a shaking-assisted vacuum filtration process to prepare large-area aligned CNT films with lateral dimensions on the order of an inch (\textcolor{blue}{Methods and Supplementary Fig.\,1}), and these CNT films have negligible defects (\textcolor{blue}{Methods and Supplementary Fig.\,2}). Further, we fabricated a two-layer twisted aligned CNTs by transferring one $\sim25$-nm-thick aligned CNT film and stacking the other with a $40^{\circ}$ rotation of alignment direction with respect to the first layer (\textcolor{blue}{Methods}). Figure\,\ref{fig:sensor_arc}b shows normalized CD spectra of a two-layer left-handed (L-twist) and a two-layer right-handed (R-twist) twisted aligned CNTs. The dominant spectral peaks occur in the ultraviolet at 267\,nm wavelength, originating from the $\pi$ resonance in CNTs~\cite{RanceEtAl2010CPL}. These peaks display opposite signs for opposite twist directions. When chiral molecules interact with twisted aligned CNTs, the amplitude of these peaks changes systematically depending on the molecule's handedness and concentration, forming the foundation of the enantioselective sensing. 

To characterize sensing performance, we developed a custom apparatus for continuous, in-situ CD spectra measurements in aqueous solutions, as depicted in Fig.\,\ref{fig:sensor_arc}c. We employed 3D printing to create a cuvette with two open windows on the sidewall, which were further sealed using broadband transparent fused silica substrates to provide water-tight confinement of solutions. In addition, we used a metallic foil to define an aperture for circularly polarized light transmission while blocking transmission through all other regions. In the middle of the cuvette, we created a narrow vertical space to securely hold sensors without movement or vibration. The cuvette's top was open to enable the easy loading and unloading of sensors and chiral molecule solutions (\textcolor{blue}{Supplementary Fig.\,3}). For each measurement of one enantiomer at one concentration, we performed four CD measurements under different configurations to remove CD responses from the linear anisotropy effect and artifacts and obtain intrinsic CD responses~\cite {DoumaniEtAl2023NC,FanEtAl2025NC} (\textcolor{blue}{Methods and Supplementary Fig.\,4}). Further, we chose D- and L-glucose chiral molecules because they exhibit no detectable CD response within our measurement wavelength and concentration ranges, as shown in Fig.\,\ref{fig:sensor_arc}d and \textcolor{blue}{Supplementary Fig.\,5}.

\subsection*{Enantioselective sensing}

\begin{figure}[H]
    \centering
    \includegraphics[width=\textwidth]{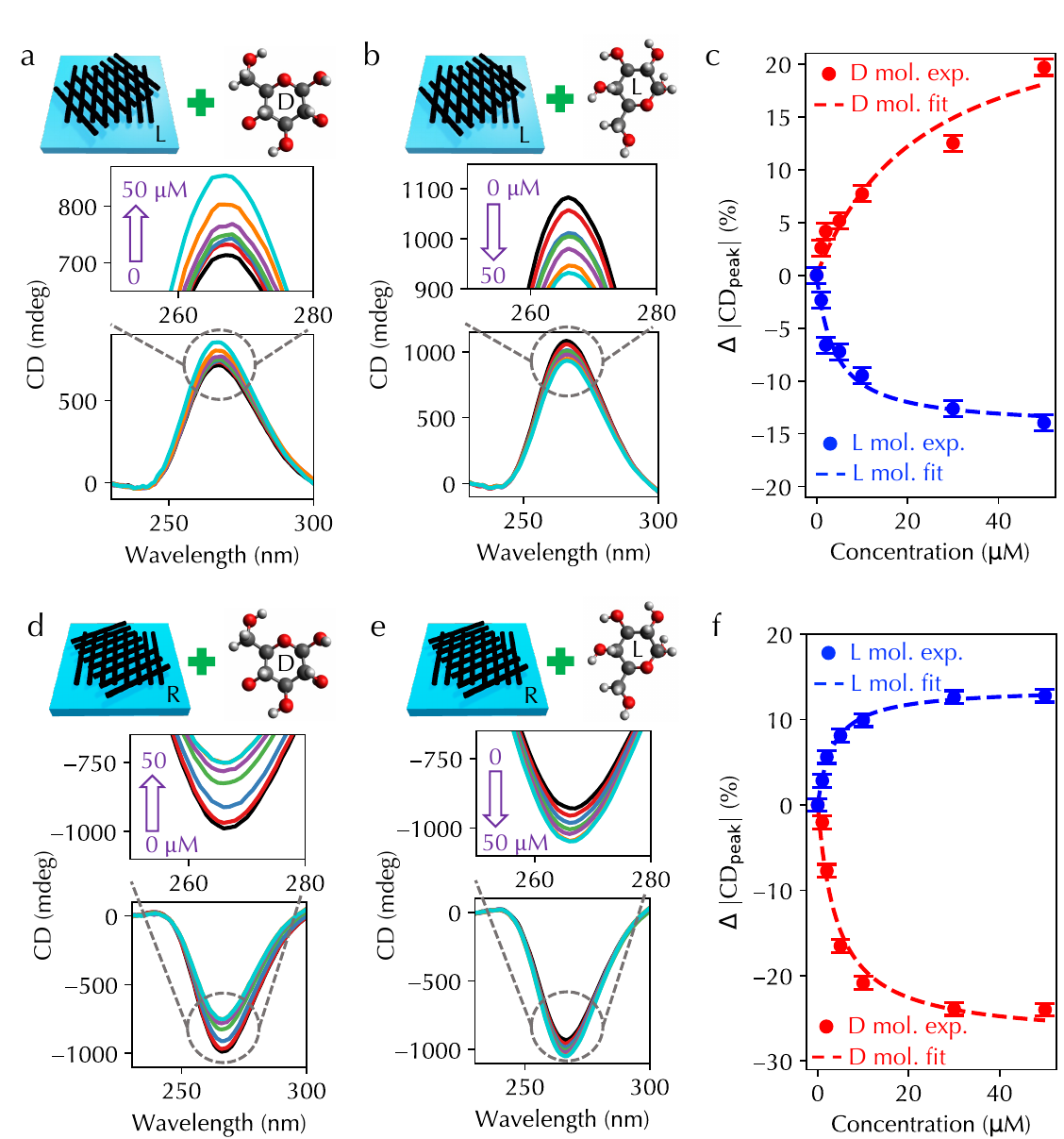}
    \vspace{-10pt}
    \caption{\textbf{Enantioselective sensing of glucose enantiomers}. Experimental CD spectra of the L-twist sensor in (a)~the D-glucose solution and (b)~the L-glucose solution with the glucose concentration from $0\,\upmu$M to $50\,\upmu$M (from black to cyan lines). (c)~Experimental (dots) and fitting (dashed lines) enantioselective sensing curves of the L-twist sensor that quantifies the percentage change of the CD peak with respect to the concentrations of D- (red color) and L-glucose solutions (blue color). (d)--(f)\,Experimental CD spectra and the enantioselective sensing curves of the R-twist sensor in the same glucose concentration range. The line and dot styles are the same with (a)--(c). }
    \label{fig:exp_results}
\end{figure}

Figure\,\ref{fig:exp_results}a and \ref{fig:exp_results}b show the experimental CD spectra of the L-twist sensor in D- and L-glucose aqueous solutions of various concentrations from 0 to 50\,$\upmu$M, respectively. Clear opposite trends of the spectral peak change, with the D-glucose solution leading to a peak amplitude increase and the L-glucose solution leading to a peak amplitude decrease, are observed. Figure\,\ref{fig:exp_results}c shows the quantified sensing curves, defined as the percentage change of the CD peak $\Delta |\textrm{CD}_\textrm{peak}(C_\textrm{mol})| = \{|\textrm{CD}_\textrm{peak}(C_\textrm{mol})| - |\textrm{CD}_\textrm{peak}(0)|\}/|\textrm{CD}_\textrm{peak}(0)|\times 100\,\%$, for the L-twist sensor in the D- and L-glucose solutions. Here, $|\textrm{CD}_\textrm{peak}(C_\textrm{mol})|$ and $|\textrm{CD}_\textrm{peak}(0)|$ are the absolute values of the CD peak or dip at a molar concentration $C_\textrm{mol}$ and in water, respectively. The highest observed $\Delta |\textrm{CD}_\textrm{peak}|$ is $\sim20$\,\% for the L-twist sensor at the $50\,\upmu$M D-glucose solution. Further, Fig.\,\ref{fig:exp_results}d and \ref{fig:exp_results}e show the CD spectra of the R-twist sensor in D- and L-glucose aqueous solutions in the same concentration range. Not only are the trends for two glucose enantiomers opposite, but also the trends of the L- and R-twist sensors for the same enantiomer are opposite, highlighting the hallmark of the enantioselective sensing. As shown in Fig.\,\ref{fig:exp_results}f, the highest observed $\Delta |\textrm{CD}_\textrm{peak}|$ is $\sim24$\,\% for the R-twist sensor at the $50\,\upmu$M D-glucose solution. 

To evaluate sensor sensitivity and LOD, we first quantified the minimum detectable $\Delta |\textrm{CD}_\textrm{peak}|$ in our four-configuration measurements. \textcolor{blue}{Supplementary Fig.\,6a} displays five repetitively measured CD spectra acquired from the same sensor, and \textcolor{blue}{Supplementary Fig.\,6b} summarizes the corresponding CD$_\textrm{peak}$ and $\Delta$CD$_\textrm{peak}$ values. From these measurements, we determined a minimum detectable $\Delta |\textrm{CD}_\textrm{peak}|<0.8\,\%$, which is shown as the error bars in the sensing curves in Fig.\,\ref{fig:exp_results}c and Fig.\,\ref{fig:exp_results}f. Further, based on the linear regions in these sensing curves, we calculated the sensitivities, defined as $\Delta|\mathrm{CD}_\mathrm{peak}|/C_\mathrm{mol}$, to be 2.6\,\%/$\upmu$M, 2.4\,\%/$\upmu$M, 2.0\,\%/$\upmu$M, and 2.8\,\%/$\upmu$M, and the minimum detectable glucose concentrations to be 0.31, 0.34, 0.39, and 0.28\,$\upmu$M for the L-twist sensor in the D-glucose solution, the L-twist sensor in the L-glucose solution, the R-twist sensor in the D-glucose solution, and the R-twist sensor in the L-glucose solution, respectively, consistently demonstrating ultrasensitive responses. The steep linear slopes and sub-$\upmu$M detection limits demonstrate an ultrasensitive, enantioselective chiroptical sensing capability that greatly outperforms prior CD glucose sensors under similar conditions~\cite{LiuEtAl2021N,XuEtAl2024AFM,KimEtAl2022N,LeeEtAl2022AFM,TromansEtAl2020CS}; see a summary in \textcolor{blue}{Supplementary Table 1}. 

Moreover, the sensing curves in Fig.\,\ref{fig:exp_results}c and Fig.\,\ref{fig:exp_results}f exhibit saturation behaviors at higher glucose concentrations, which is a clear signature of the adsorption saturation of glucose molecules on the CNT sidewall and can be modeled using a Langmuir isotherm
\begin{align}
\theta = \frac{K_\mathrm{mol}C_\mathrm{mol}}{1+K_\mathrm{mol}C_\mathrm{mol}},
\label{main_eq:langmuir}
\end{align}
where $\theta$ is the fractional molecule coverage and $K_\mathrm{mol}$ is the adsorption equilibrium constant in a unit of inverse concentration corresponding to $C_\mathrm{mol}$. Hence, we hypothesize that $\Delta|\textrm{CD}_\textrm{peak}(C_\mathrm{mol})|$ follows a single governing equation
\begin{align}
\Delta |\textrm{CD}_\textrm{peak}| = h_\textrm{twist}h_\textrm{mol} \Delta \textrm{CD}_\textrm{trans} \theta =  h_\textrm{twist}h_\textrm{mol} \Delta \textrm{CD}_\textrm{trans} \frac{K_\mathrm{mol}C_\mathrm{mol}}{1+K_\mathrm{mol}C_\mathrm{mol}}.
\label{main_eq:govern_eqs}
\end{align}
Here, $h_\textrm{twist} = \pm1$ and $h_\textrm{mol} = \pm1$ describe the handednesses of the twist sensor and the glucose enantiomer, respectively. Specifically, $h_\textrm{twist} = 1$ ($h_\textrm{twist} = -1$) means the L-twist (R-twist) sensor, and $h_\textrm{mol} = 1$ ($h_\textrm{mol} = -1$) means the D-glucose (L-glucose) solution. $\Delta \textrm{CD}_\textrm{trans}$ is the sensing transduction strength that maps a microscopic molecule coverage $\theta$ to the experimentally measurable macroscopic $\Delta \textrm{CD}_\textrm{peak}$ response. We fit experimental results well using Eq.\,\ref{main_eq:govern_eqs} with $\Delta \textrm{CD}_\textrm{trans}$ and $K_\textrm{mol}$ as fitting parameters, as shown in the dashed lines of Fig.\,\ref{fig:exp_results}c and Fig.\,\ref{fig:exp_results}f. 

The extracted $\Delta \textrm{CD}_\textrm{trans}$ and $K_\textrm{mol}$ values for the four cases, the L-twist sensor in the D-glucose solution, the L-twist sensor in the L-glucose solution, the R-twist sensor in the D-glucose solution, and the R-twist sensor in the L-glucose solution, are 26.7 and $4.2\times10^4\,\mathrm{M}^{-1}$, 14.4 and $2.5\times10^5\,\mathrm{M}^{-1}$, 27.4 and $2.3\times10^5\,\mathrm{M}^{-1}$, and 13.7 and $3.0\times10^5\,\mathrm{M}^{-1}$, respectively. Despite a relatively lower $K_\textrm{mol}$ value in the first case compared to others, which we attribute to a slight shift of the measurement sample location, $K_\textrm{mol}$ values are well above $10^{5}\,\mathrm{M}^{-1}$. Notably, the extracted $K_\textrm{mol}$ values are substantially (i.e., at least one order of magnitude) larger than those reported in various platforms, such as glucose oxidase ($\sim110\,\mathrm{M^{-1}}$)~\cite{YangEtAl2016SAC}, alkali-activated microporous carbon ($\sim220\,\mathrm{M^{-1}}$)~\cite{YabushitaEtAl2014C}, diboronic-acid receptors ($\sim2000 - 6500\,\mathrm{M^{-1}}$)~\cite{NanEtAl2023B}, glucose oxidase-functionalized CNTs ($\sim 830\,\mathrm{M^{-1}}$)~\cite{BaroneEtAl2005NM}, Concanavalin A ($\sim500\,\mathrm{M^{-1}}$)~\cite{TooneEtAl1994COSB}, and some synthetic lectin systems ($\sim250\,\mathrm{M^{-1}}$)~\cite{RiosEtAl2017CS}. Further, our values exceed those of high-affinity nanomaterials, proteins, and receptors, such as graphene quantum dots ($\sim4\times10^4\,\mathrm{M^{-1}}$)~\cite{RajmaneEtAl2025SR}, some glucose-binding protein ($\sim7700\,\mathrm{M^{-1}}$)~\cite{VeetilEtAl2010BB} and biomimetic receptor ($\sim1.8\times10^{4}\,\mathrm{M^{-1}}$)~\cite{TromansEtAl2019NC}, while they are still smaller than some special glucose-binding protein ($\sim1.3\times10^7\,\mathrm{M^{-1}}$)~\cite{CuneoEtAl2006JMB}; see a summary in \textcolor{blue}{Supplementary Table\,2}. The ultrahigh $K_\textrm{mol}$ values promote near-surface enrichment of chiral molecules even at low bulk concentration, thereby enabling substantial adsorption-driven chiroptical transduction in the sub-$\upmu$M regime and providing the physical basis for the ultrasensitive detection.

Further, the extracted $\Delta \textrm{CD}_\textrm{trans}$ in the D-glucose solution is approximately twice that in the L-glucose solution, indicating different chiroptical transduction per adsorption site for two enantiomers. A plausible explanation is that the adsorption-induced perturbation of glucose molecules that couples to the CD spectra of twisted aligned CNTs is dominated by an axially projected chiroptical response along the CNT alignment direction~\cite{AllenmarkEtAl2003C}. Hence, the spectral change depends not only on the number of adsorbed molecules but also on the orientation of glucose at the interface. One possible origin of the observed asymmetry is trace residue of the chiral surfactant sodium deoxycholate (DOC) used during CNT dispersion and film processing. Although DOC itself and any residual DOC in CNT films do not exhibit any measurable CD signals~\cite{DoumaniEtAl2023NC}, even trace DOC could template interfacial hydrogen-bonding and hydration networks~\cite{XiongEtAl2024COCIS}, thereby biasing the interfacial orientations of D- and L-glucose around achiral CNT sidewalls.

In addition to glucose, we demonstrate that our sensor platform can generalize to other chiral molecules by sensing alanine, which is a representative chiral amino acid. \textcolor{blue}{Supplementary Fig.\,7}a shows the experimental CD spectra acquired from the R-twist sensor in the L-alanine solution, and \textcolor{blue}{Supplementary Fig.\,7}b summarizes the corresponding sensing curve. Fitting these data using the Langmuir isotherm yields $\Delta \textrm{CD}_\textrm{trans} = 8.5$ and $K_\textrm{mol} = 7.9\times 10^{5}\,\mathrm{M^{-1}}$, again suggesting ultrahigh interfacial affinity and efficient adsorption-driven chiroptical transduction. Consistently, the extracted sensitivity and LOD are 3.4\,\%/$\upmu$M and 0.24\,$\upmu$M, respectively, comparable to those obtained for L-glucose and exhibiting the same trend of CD peak change. Together, these results confirm that our sensors are not specific to glucose but provide a generic, adsorption-mediated chiroptical sensing strategy that can be extended to a variety of chiral molecules, including carbohydrates and amino acids.

\subsection*{Molecular dynamics simulation}

\begin{figure}[H]
    \centering
    \includegraphics[width=\textwidth]{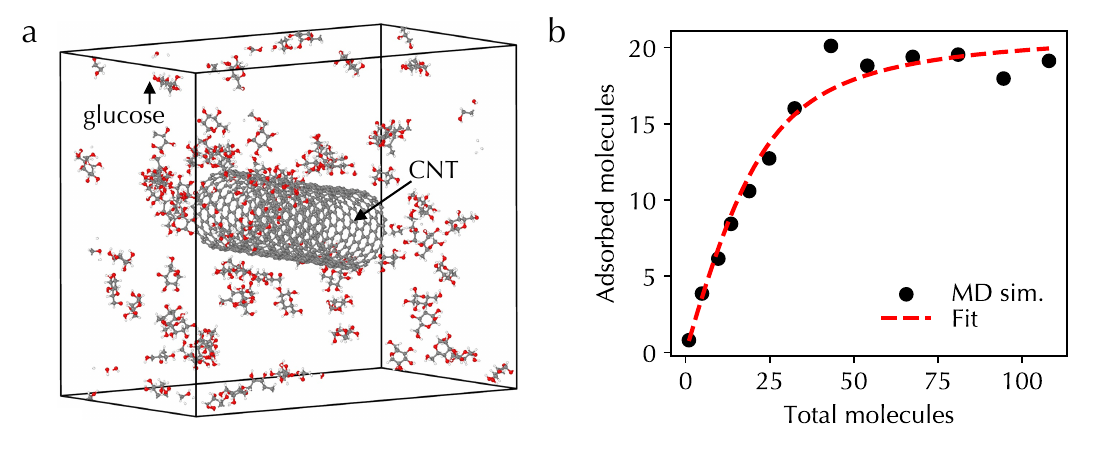}
    \vspace{-15pt}
    \caption{\textbf{MD simulation}. (a)~Illustration of the MD simulation setup, including an achiral CNT surrounded by D-glucose and water molecules (not included for clarity). (b)~Simulation (black dots) and fitting (red dashed line) number of adsorbed glucose molecules with respect to the total number of glucose molecules in the system.}
    \label{fig:md_sim}
\end{figure}

To further validate the adsorption process, we carried out the equilibrium MD simulations of an achiral CNT surrounded by various numbers of D-glucose molecules and a fixed number of water molecules, as illustrated in Fig.\,\ref{fig:md_sim}a. We chose an achiral zigzag CNT with a diameter close to the average diameter of CNTs used in our experiments; see \textcolor{blue}{Methods} for details of the MD simulation setup. Figure\,\ref{fig:md_sim}b shows the number of adsorbed glucose molecules with respect to the number of total molecules in the system, revealing a clear saturation behavior. We then developed a closed-system Langmuir isotherm to fit simulation results and obtained a molecule-count-based adsorption equilibrium constant; see \textcolor{blue}{Methods} for detailed expressions.   

To compare with experiments, we converted the adsorption equilibrium constant from a molecule-count basis to an equivalent molar-concentration form. Note that directly simulating the experimental conditions at low glucose concentrations requires an enormous simulation domain and thus prohibitive computational resources to include enough glucose molecules for meaningful statistics. Instead, our simulations employed a substantially smaller simulation domain volume, noting that adsorption is primarily influenced by the local environment within the periodic simulation domain, while water molecules beyond this region only have a minor impact. Accordingly, we introduced an effectively larger conversion volume, $V_\mathrm{cv}$, to bridge the molecule-count and molar-concentration descriptions by mapping glucose molecule numbers to their corresponding molar concentrations, $C_\mathrm{mol}$. This mapping was established by assuming that the coverage $\theta$ computed from molecule counts was the same as that in Eq.\,\ref{main_eq:langmuir}. Using the resulting $V_\mathrm{cv}$, we converted the molecule-count-based equilibrium constant to $K_\mathrm{mol}$, obtaining $K_\mathrm{mol}\sim2.4\times10^5\,\mathrm{M^{-1}}$ in excellent agreement with experiments; see \textcolor{blue}{Methods} for detailed conversion expressions. 

We further repeated the MD simulations using a larger periodic simulation domain, and \textcolor{blue}{Supplementary Fig.\,8} shows simulation results. Applying the same closed-system Langmuir fitting and conversion procedures yielded $K_\mathrm{mol}\sim2.5\times10^5\,\mathrm{M^{-1}}$, confirming the self-consistency of the analysis and showing that $K_\mathrm{mol}$ is effectively invariant with respect to the simulation domain volume.

\subsection*{Optical simulation}

\begin{figure}[H]
    \centering
    \includegraphics[width=\textwidth]{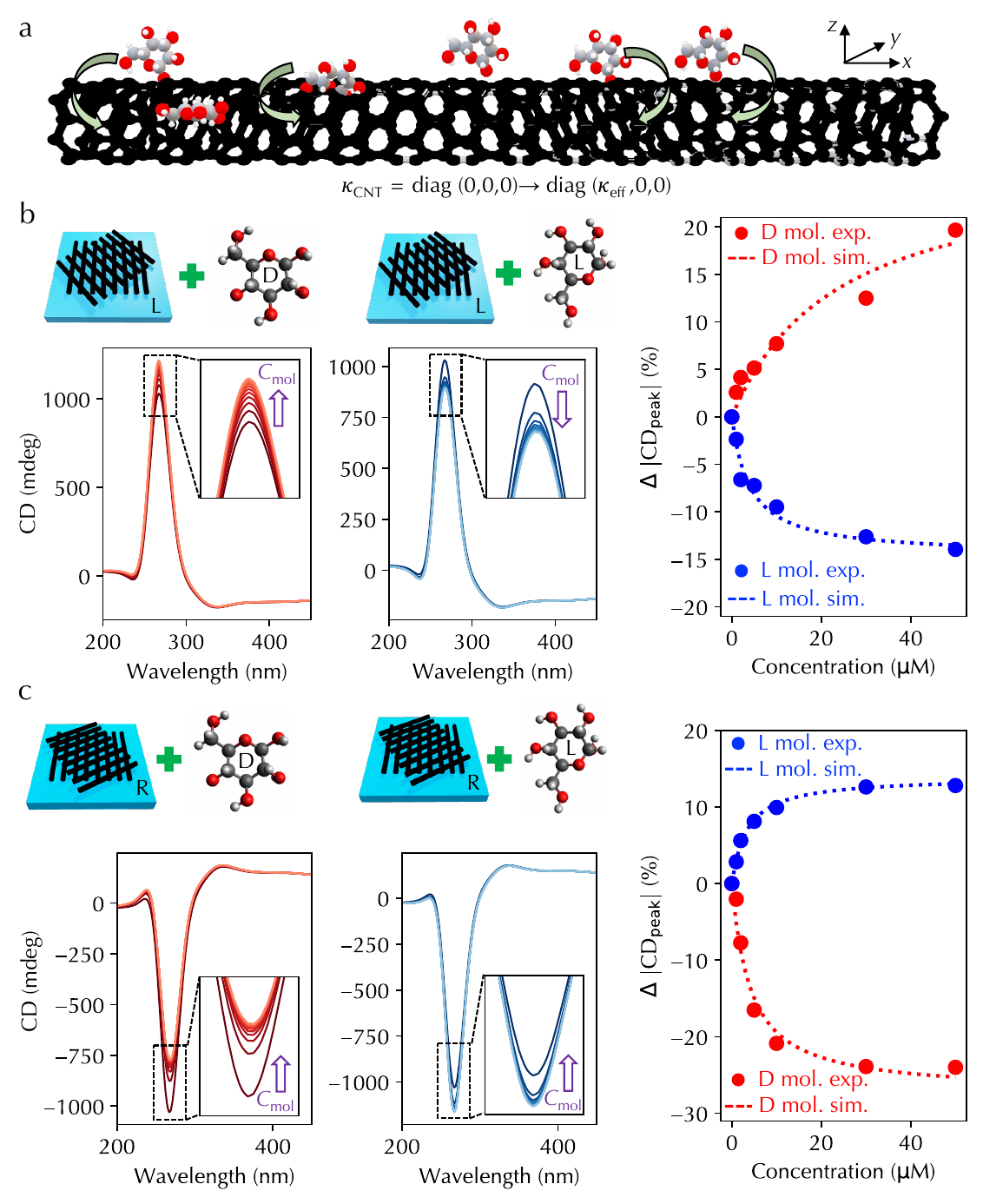}
    \vspace{-15pt}
    \caption{\textbf{Optical simulation using chiral transfer matrix method}. (a)~Illustration of the adsorption of chiral molecules on a CNT, leading to a nonzero chirality parameter along the CNT axis. (b)~Simulation CD spectra for the L-twist sensor in D- (dark to light red lines) and L-glucose solutions (dark to light blue lines), and corresponding experimental (dots) and simulation (lines) enantioselective sensing curves under various concentrations of D- (red color) and L-glucose (blue color) solutions from $0\,\upmu$M to $50\,\upmu$M. (c)~Simulation CD spectra, and experimental and simulation enantioselective sensing curves for the R-twist sensor. The line and dot styles are the same as (b).}
    \label{fig:optical_sim}
\end{figure}

To interpret the experimental CD spectra and enantioselective sensing curves and to extract the underlying material parameters, we developed the cTMM for general bianisotropic media and performed optical simulations. As illustrated in Fig.\,\ref{fig:optical_sim}a, the individual CNTs used in sensors do not exhibit any chiroptical responses and thus the chirality parameter, $\kappa_\mathrm{CNT}$, is zero in all directions. Hence, the chiroptical responses of twisted aligned CNT stacks in the absence of adsorbates can be simulated using the transfer matrix method assuming standard electromagnetic constitutive relations without the coupling between electric and magnetic fields~\cite{DoumaniEtAl2023NC,FanEtAl2025NC}. In contrast, the adsorption of chiral molecules on CNTs creates an effective composite with a breaking symmetry, inducing a nonzero $\kappa_\mathrm{CNT}$~\cite{AllenmarkEtAl2003C}. Note that perturbations to the dielectric functions along and perpendicular to the CNT axis alone cannot account for the experimentally observed enantioselective trends, because enantiomers produce the dielectric changes with the same sign. Hence, reproducing opposite-direction sensing curves requires an optical quantity that changes signs with handedness, such as the chirality parameter. Accordingly, the constitutive relations and wave equations in the cTMM require the incorporation of the coupled electric and magnetic field components~\cite{LindellEtAl1994,FanEtAl2025C}. We implemented a cTMM solver using the PyTorch framework to efficiently and fast compute reflection, transmission, and absorption spectra for arbitrary multilayer stacks under any linear or circular polarizations (\textcolor{blue}{Methods}). 

Although it is well established to model anisotropic dielectric functions of aligned CNTs and extract them by fitting experimental spectra (\textcolor{blue}{Methods and Supplementary Fig.\,9})~\cite{DoumaniEtAl2023NC,FanEtAl2025NC}, the modeling of the chiral-molecule-induced chirality parameter in an anisotropic system has, to our knowledge, not been explored; here we developed such a model for the first time. Specifically, because of the 1D geometry in aligned CNTs, it is reasonable to assume that the effective chirality parameter tensor of the chiral-molecule-CNT composite inherits the same anisotropy with a nonzero chirality parameter component, $\kappa_\mathrm{eff}$, along the CNT axis and zero components in other directions; see Fig.\,\ref{fig:optical_sim}a. Further, we modeled $\kappa_\mathrm{eff}$ to be dispersive using a single-frequency Condon model, and the amplitude of $\kappa_\mathrm{eff}$ to be proportional to the coverage $\theta$ with a proportional scaling coefficient, $F_\mathrm{s}$, the same for both D- and L-glucose molecules. We applied the obtained fitting parameters $\Delta\mathrm{CD}_\mathrm{trans}$ and $K_\mathrm{mol}$ from the experiments (Fig.\,\ref{fig:exp_results}) to the $\kappa_\mathrm{eff}$ expression (\textcolor{blue}{Methods}). 

Figure\,\ref{fig:optical_sim}b and Fig.\,\ref{fig:optical_sim}c show the simulation CD spectra and corresponding enantioselective sensing curves of the L-twist and R-twist sensors, which have the same structural parameters as the experiments with two 25-nm-thick aligned CNT films and a rotation angle of $40^\circ$, in various concentrations of D- and L-glucose solutions from $0\,\upmu$M to $50\,\upmu$M. All simulation results show excellent agreement with experiments, validating our optical anisotropic modeling of the chiral-molecule-CNT composite and the cTMM solver. \textcolor{blue}{Supplementary Fig.\,10} displays the extracted real and imaginary parts of $\kappa_\mathrm{eff}$ for the composite with D- and L-glucose under various concentrations. A small $\kappa_\mathrm{eff}$ on the order of $10^{-3}$ can lead to substantial CD spectra change, highlighting the ultrasensitive performance in our sensors. 

\subsection*{Programmability}

\begin{figure}[hbt]
    \centering
    \includegraphics[width=\textwidth]{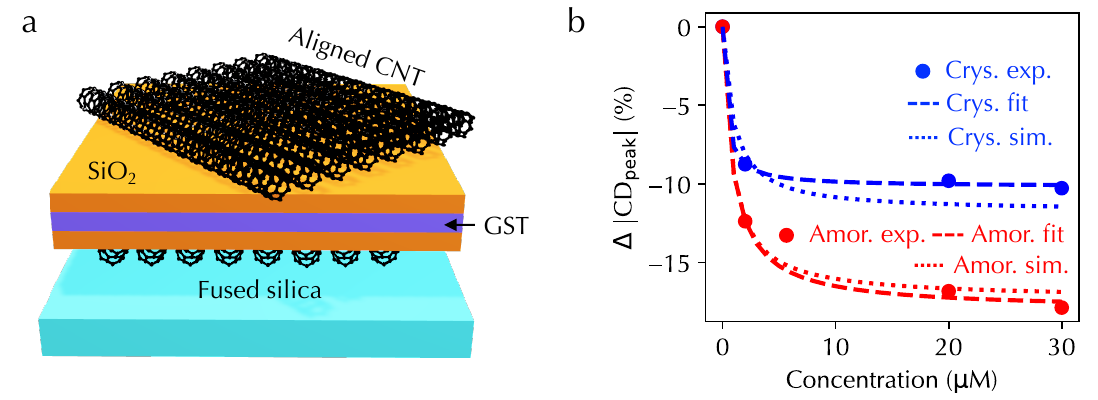}
    \vspace{-15pt}
    \caption{\textbf{Programmable sensor}. (a)~Schematic diagram of the programmable sensor by incorporating a GST film between twisted aligned CNTs. (b)~Experimental (dots), fitting (dashed lines), and simulation (dotted lines) programmable sensing curves under amorphous (red color) and crystalline phases (blue color) of the GST, respectively.}
    \label{fig:tune_sensing}
\end{figure}

Moreover, the layered sensor architecture offers the flexibility of incorporating other materials to bring new functionality to sensors. Figure\,\ref{fig:tune_sensing}a illustrates a programmable sensor by inserting a germanium-antimony-tellurium (Ge$_2$Sb$_2$Te$_5$ or GST) thin film between two layers of twisted aligned CNTs. GST is one type of nonvolatile PCM, which can have reversible phase transitions between amorphous and crystalline phases under external excitation and exhibit a substantially large change of refractive indices or dielectric functions during the phase transition~\cite{HuangEtAl2025ARMR}. As a result, both the sensor’s chiroptical response~\cite{FanEtAl2025NC} and its chiroptical transduction upon the interaction with chiral molecules can be programmed. 

We fabricated an L-twisted sensor incorporating an 8-nm-thick GST film sandwiched between a 155-nm-thick SiO$_2$ layer and a 55-nm-thick SiO$_2$ layer, and experimentally characterized the sensing curves for L-glucose under the amorphous and crystalline GST phases, as shown in Fig.\,\ref{fig:tune_sensing}b. Fitting both curves with the Langmuir isotherm yields $\Delta\mathrm{CD}_\mathrm{trans} = 18.0$ and $K_\mathrm{mol} = 1.1\times10^6\,\mathrm{M^{-1}}$ in the amorphous GST state, and $\Delta\mathrm{CD}_\mathrm{trans} =10.2$ and $K_\mathrm{mol} = 3.0\times10^6\,\mathrm{M^{-1}}$ in the crystalline GST state, demonstrating that switching the GST phase dynamically programs the chiroptical transduction amplitude and adsorption affinity, and thus the overall sensing curve, without altering the sensor geometry. Moreover, we employed optical constants of GST films~\cite{ZhengEtAl2018OME} and cTMM simulations to reproduce both sensing curves with good agreement, further validating the material modeling and the solver.

\section*{Conclusion}

Taken together, we demonstrated a conceptually new enantioselective sensing paradigm that unifies molecule-level chirality induction with device-level optical transduction in a wafer-scale, programmable heterostructure built from twisted stacks of adsorption-mediated aligned CNT films and GST films. By leveraging adsorption-driven formation of chiroptically active CNT-molecule composites and amplifying their responses through self-assembled macroscopic alignment, deterministic rotational stacking, and strong ultraviolet CNT chiroptical resonances, our platform converted weak molecular chiroptical responses into large, readily measurable CD signals without nanofabricated hotspots or differential left- and right-handed device protocols. Using CD-silent glucose as a stringent benchmark and extending to chiral amino acids, we realized continuous, in situ sensing that simultaneously captures enantiomer handedness and concentration with a single device, achieving sub-µM LOD enabled by a large adsorption equilibrium constant ($>10^5\,\mathrm{M^{-1}}$). Further, we employed MD simulations to corroborate molecular adsorption on CNTs, and cTMM-based optical simulations to reproduce experimentally obtained CD spectra and enantioselective sensing behavior, together providing a unified mechanistic framework that links experiment and simulation. We further demonstrated the sensor programmability by integrating and tuning a GST film between twisted CNT stacks, providing a practical route toward reconfigurable sensing performance. 

Looking forward, this architecture opens several directions for advancing chiroptical sensing from laboratory demonstrations to deployable, real-time monitors in complex aqueous environments. From the material perspective, tailoring adsorption chemistry, quantum-engineering CNT optical transitions, employing CNT intrinsic chiralities, and extending to other 1D nanomaterials~\cite{XuEtAl2022CEJ,XiangEtAl2020S} could extend dynamic range, suppress interferents, and generalize detection to broader classes of clinically and industrially relevant chiral molecules. From the device perspective, the cTMM solver built on a differentiable PyTorch framework can automatically design the programmable sensor to meet application-specific requirements. Given a target chiral molecule and desired performance metrics, such as wavelength window, sensitivity, dynamic range, and constraints on stack thickness or materials, the solver can optimize the multilayer architecture and tuning states to deliver the required enantioselective response on demand.

\section*{Methods}

\smallskip
\noindent\textbf{Preparation of wafer-scale aligned CNT films} -- Shaking-assisted vacuum filtration (SAVF) was used to prepare wafer-scale aligned CNT films. The filtration process started with the preparation of a well-dispersed CNT aqueous suspension. Specifically, 8\,mg arc-discharge-synthesized CNT powder with an average diameter of 1.4\,nm (Carbon Solutions, Inc. P2 product) was mixed with a 20\,mL $0.5\%$ w/w DOC solution, and then the mixture was sonicated using an ultrasonic tip horn sonicator (QSonica Q125) for 45 minutes at an output power of 21\,W. The sonicated suspension was further purified using ultracentrifugation at 247,104\,g for two hours to remove large bundles, and the supernatant was collected. For the vacuum filtration process, a 1-inch filtration system (MilliporeSigma) was mounted on a linear shaker (Scilogex SCI-L180-Pro LCD digital linear shaker); see \textcolor{blue}{Supplementary Fig.\,1}. A diluted supernatant suspension was poured into the filtration system through a funnel with a square-shaped cross-section and a 100-nm-pore-size polycarbonate filter membrane (Whatman Nuclepore Track-Etched polycarbonate hydrophilic membranes, MilliporeSigma) for CNT alignment, which was determined by the linear shaking direction~\cite{FanEtAl2025NC}. Specifically, the linear shaker shook the filtration system at 200\,RPM during the first 15\,minutes of the process, and the rest of the process continued without shaking. 

\smallskip
\noindent\textbf{Raman spectroscopy} -- The Raman spectroscopy was carried out to estimate the defects in CNT films using an iHR 550 Horiba Raman spectrometer equipped with a CCD camera and a 4\,mW excitation laser at the wavelength of 266\,nm (CryLas FQCW266-10), which was on resonance with the $\pi$-peak of CNT absorption spectra. The characteristic distance between defects in a CNT, $L_a$, can be calculated by $L_a = \frac{560\textrm{(nm)}}{E_l^4}(I_G/I_D)$, where $E_l$ is the laser excitation energy in units of eV and $I_G/I_D$ is the Raman intensity ratio of the G peak to the D peak. \textcolor{blue}{Supplementary Fig.\,2} displays a Raman spectrum of a prepared aligned CNT film and no clear D peak was observed, suggesting a low and negligible defect density in our film. 

\smallskip
\noindent\textbf{Sensor fabrication} --  One aligned CNT film on polycarbonate filter membranes prepared through the SAVF process was first transferred onto broadband transparent fused silica substrates using a wet transfer method~\cite{HeEtAl2016NN}. Specifically, a small droplet of water was placed on the substrate, and the CNT film on the polycarbonate filter membrane was placed with the CNT side in contact with the wet substrate and the polycarbonate side on top. Once the water between the CNT film and the substrate evaporated, the top polycarbonate layer was removed by immersing the sample in a chloroform solution. The sample was finally cleaned with isopropanol. Similarly, another aligned CNT film was transferred onto the first one with a rotated alignment direction to create twisted aligned CNTs. To fabricate a programmable sensor, SiO$_{2}$ films were deposited using a Denton Discovery 18 Sputtering System at an argon pressure of 6\,mTorr with a power setting of 100\,W. A GST film was deposited using the same system at an argon pressure of 4.5\,mTorr and a power setting of 35\,W. To induce the phase transition in the GST film, the heterostructure was placed on a hotplate at a set temperature and stayed for a few minutes for a complete phase transition. 

\smallskip
\noindent\textbf{Sensor characterization} -- The first type of chiral molecule used for characterizing sensors was glucose. To prepare glucose solutions of various concentrations, a 10 mM glucose stock solution was first prepared by weighing 36\,mg of glucose powder and dissolving it in approximately 15\,mL of deionized water. The solution was then transferred quantitatively to a 20\,mL volumetric flask and diluted to the mark with deionized water to obtain a final concentration of 10\,mM. The stock was mixed thoroughly and ultrasonicated for 10\,minutes. The solutions of other concentrations were prepared by serial dilution of this stock. For example, a 2\,mM glucose solution was obtained by mixing 4\,mL of the 10\,mM stock with 16\,mL of deionized water to a final volume of 20\,mL; a 0.5\,mM glucose solution was prepared by diluting 5\,mL of the 2\,mM solution with 15\,mL of deionized water to 20\,mL; and a 50\,$\upmu$M glucose solution was prepared by mixing 2\,mL of the 0.5\,mM solution with 18\,mL of deionized water to a final volume of 20\,mL. The second type of chiral molecule used for characterization was the alanine solution, which was prepared in a similar way to the glucose solution. 

CD spectra were captured using a standard JASCO J-815 CD spectrometer in a wavelength range from $200-900\,$nm and our custom cuvette with an optical beam diameter of 2\,mm. Although the optical path length of the cuvette was 8\,mm, the actual interaction between chiral molecules and a sensor occurred in a length of sensor thickness, which was $\sim50\,$nm for CNT-only sensors and $\sim250\,$nm for CNT-PCM-SiO$_2$ heterostructures. Because of the anisotropic structure in our sensor architecture, the measured CD spectra contained the contributions from intrinsic structure-induced chirality, linear-anisotropy-induced effect, and system artifacts. To single out the CD response from the intrinsic structure-induced chirality, four CD spectra were captured for each measurement of one enantiomer under one concentration by flipping and rotating the sensor, as illustrated in \textcolor{blue}{Supplementary Fig.\,4}, and the average of these four spectra represented the intrinsic CD spectra~\cite{DoumaniEtAl2023NC,FanEtAl2025NC}. Before each four-configuration measurement, the sensor was immersed in the solution for a few minutes to reach a stable state. The typical diffusion constants, $D_\textrm{diff}$, of small organic molecules, such as glucose and alanine, are from $10^{-6}$ to $10^{-5}\,$cm$^{2}$/s and the CNT thickness in sensors, $d_\textrm{sensor}$, is generally $<100\,$nm. Hence, the diffusion time, $\tau_\textrm{diff}$, needed to reach a stable condition can be estimated as $\tau_\textrm{diff} = d_\textrm{sensor}^2/D_\textrm{diff} < 0.1\,$ms, which is much smaller than the experimental waiting time. Further, after one complete four-configuration measurement, the sensor was rinsed thoroughly using deionized water and dried for the next measurement.

\smallskip
\noindent\textbf{MD simulation and closed-system Langmuir adsorption isotherm} -- Equilibrium MD simulations were performed using LAMMPS~\cite{ThompsonEtAl2022CPC}, and trajectory visualization was carried out with OVITO~\cite{StukowskiEtAl2009MSMSE}. An (18,0) zigzag CNT was positioned at the center of the simulation box and surrounded by water and glucose molecules. The CNT axis was aligned along the $z$-direction of the simulation domain. The CNT had a length of 4.2\,nm and a diameter of 1.4\,nm. Water molecules were initially generated on a simple cubic lattice with a lattice constant of 0.310\,nm. The simulation box dimensions were $4.9\times4.5\times4.2$\,nm, containing a total of 3024 water molecules. Glucose molecules were randomly distributed within the water. To prevent atomic overlap, glucose molecules were excluded from the interior of the CNT as well as from a 0.5\,nm region surrounding the CNT surface. The CHARMM36m force field was employed for glucose molecules, TIPS3P was used to model water, and the CNT was described using the Tersoff potential. The SHAKE algorithm was used to constrain the bond lengths and bond angles in water molecules, allowing the use of a larger integration time step. Non-bonded interactions between CNT, water, and glucose were modeled using a 12-6 Lennard-Jones (LJ) potential~\cite{HummerEtAl2001N,ZhaoEtAl2007JACS,ZhangEtAl2019JPC,ChenEtAl2024L}. The LJ parameters for CNT carbon atoms were $\epsilon_\mathrm{LJ} = 0.086\,\mathrm{kcal\,mol^{-1}}$ and $\sigma_\mathrm{LJ} = 0.34~\mathrm{nm}$.

The two ends of the CNT were fixed throughout the simulations. Periodic boundary conditions were applied in all three spatial directions. Prior to the production run, the initial configuration was energy-minimized using the conjugate gradient algorithm. The system was then simulated for 4\,ns in the canonical (constant number of particles, constant volume, and constant temperature; NVT) ensemble with a time step of 0.5\,fs. Nos\'e Hoover thermostat was applied to set the temperature at 298\,K. After the system reached equilibrium, atomic trajectories were recorded every 5\,ps over a subsequent duration of 4\,ns for analysis. A glucose molecule was considered adsorbed onto the CNT if the center of mass of the molecule was within a specified cutoff distance from the CNT surface. The cutoff distance was determined from the radial distribution function (RDF) between the CNT and glucose molecules, calculated using the centers of mass of glucose molecules. The radial distance corresponding to the first peak of the RDF was taken as the adsorption cutoff. The average value was $\sim0.5$\,nm. The number of adsorbed glucose molecules was evaluated for each saved time frame and averaged over 4\,ns. 

A modified Langmuir adsorption isotherm for a closed system, which contained a fixed total number of glucose molecules that could adsorb onto the CNT surface in a finite volume, was developed to fit MD simulation results~\cite{ZangiEtAl2024L}. A standard Langmuir isotherm assumes a fixed free concentration $C$ by a reservoir and is expressed as 
\begin{equation}
    \theta = \frac{N_{\mathrm{ads}}}{N_{\mathrm{site}}} = \frac{KC}{1+KC},
    \label{eq:langmuir_Nads}
\end{equation}
where $\theta$ is the fractional coverage ($0 \le \theta \le 1$), $N_{\mathrm{ads}}$ is the number of molecules adsorbed at equilibrium, $N_{\mathrm{site}}$ is the maximum possible number of adsorbable molecules (i.e., site capacity), $K$ is the affinity or equilibrium constant, and $C$ is calculated in terms of counts as $N_{\mathrm{free}}/V$ with $N_{\mathrm{free}}$ as the number of free molecules at equilibrium and $V$ as the volume. 

However, in a closed system, adsorption depletes the free pool. Hence, $C$ is not fixed and becomes
\begin{equation}
    C = \frac{N_{\mathrm{tot}} - N_{\mathrm{ads}}}{V}.
    \label{eq:C_closed}
\end{equation}
Substituting Eq.~\eqref{eq:C_closed} into Eq.~\eqref{eq:langmuir_Nads} yields an implicit closed-system equation
\begin{equation}
  \frac{N_{\mathrm{ads}}}{N_{\mathrm{site}}} = 
  \frac{K\left(\frac{N_{\mathrm{tot}} - N_{\mathrm{ads}}}{V}\right)
  }{
    1 + K\left(\frac{N_{\mathrm{tot}} - N_{\mathrm{ads}}}{V}\right)
  }.
  \label{eq:closed_implicit}
\end{equation}
It is convenient to define a parameter $a$ as $K/V$ with a unit of inverse molecule count. Hence, the closed-system Langmuir equation becomes
\begin{equation}
  \frac{N_{\mathrm{ads}}}{N_{\mathrm{site}}} =  \frac{a\,(N_{\mathrm{tot}} - N_{\mathrm{ads}})}{1 + a\,(N_{\mathrm{tot}} - N_{\mathrm{ads}})},
  \label{eq:closed_a}
\end{equation}
which can be solved analytically for $N_{\mathrm{ads}}$ as a function of $N_{\mathrm{tot}}$
\begin{equation}
  N_{\mathrm{ads}}(N_{\mathrm{tot}}) =
  \frac{
    \left(N_{\mathrm{site}} + N_{\mathrm{tot}} + \frac{1}{a}\right)
    -
    \sqrt{
      \left(N_{\mathrm{site}} + N_{\mathrm{tot}} + \frac{1}{a}\right)^2 - 4N_s N_{\mathrm{tot}}
    }
  }{2}.
  \label{eq:explicit_solution}
\end{equation}
The minus sign is in front of the square root, ensuring $0\le N_{\mathrm{ads}}\le \min(N_{\mathrm{site}}, N_{\mathrm{tot}})$.

Equation\,\eqref{eq:explicit_solution} was utilized to fit MD simulation results with two fitting parameters, $N_\mathrm{site}$ and $a$, which were $21.36\,\pm\,0.92$ and $0.16\,\pm\,0.04$. Further, the dimensionless  product, $KC$, can be obtained from
\begin{equation}
  K C = K\frac{N_{\mathrm{free}}}{V} = a\,N_{\mathrm{free}}.
  \label{eq:KC_lumped}
\end{equation}

In experiments, the concentration, $C_\mathrm{mol}$, is expressed in a unit of M and the corresponding adsorption equilibrium constant $K_\mathrm{mol}$ has a unit of (M$^{-1}$). However, directly performing MD simulations at the low glucose concentrations (up to $50\,\upmu$M) used in experiments requires an unrealistically large simulation domain to include a statistically meaningful number of chiral molecules. Indeed, water molecules outside the MD simulation domain only have minor effects, and the adsorption behavior is primarily governed by the environment within the periodic simulation cell. Hence, $K$ can be converted to $K_\mathrm{mol}$ by choosing an appropriate volume, $V_\mathrm{cv}$, to convert $N_\mathrm{tot}$ to $C_\mathrm{mol}$. This conversion was done by assuming the same coverage $\theta$. Specifically, at the largest number of adsorbed molecules, $N_\mathrm{ads, max}$, the largest $\theta$, $\theta_\mathrm{max}$, in MD simulations was calculated using $N_\mathrm{ads, max}/N_\mathrm{site}$ following Eq.\,\ref{eq:langmuir_Nads}. The obtained $\theta_\mathrm{max}$ was further used to calculate $C_\mathrm{mol, max}$ following the expression $\theta_\mathrm{max}/{K_\mathrm{mol, avg}(1-\theta_\mathrm{max})}$ from Eq.\,\ref{main_eq:langmuir}. Thus, the conversion between $N_\mathrm{ads, max}$ and $C_\mathrm{mol}$ was expressed as
\begin{equation}
    C_\mathrm{mol, max} = \frac{N_\mathrm{ads, max}}{K_\mathrm{mol, avg}(N_\mathrm{site}-N_\mathrm{ads, max})},
\end{equation}
where $K_\mathrm{mol, avg}=2\times10^5\,\mathrm{M^{-1}}$ is the equilibrium constant averaged over four experimental sensing curves in Fig.\,\ref{fig:exp_results}. Thus, $V_\mathrm{cv}$ was calculated as
\begin{equation}
    V_\mathrm{cv} = \frac{N_\mathrm{tot,max}}{C_\mathrm{mol,max}N_A},
\end{equation}
where $N_A$ is Avogadro's number. Specifically for MD simulations in Fig.\,\ref{fig:md_sim}, the fitting $N_\mathrm{ads,max}\sim 20$, $N_\mathrm{tot,max}\sim108$, $C_\mathrm{mol,max}\sim70\,\upmu$M, and $V_\mathrm{cv}\sim2.6\times10^6\,\mathrm{nm^3}$. Further, the dimensionless product can also be expressed as
\begin{equation}
  KC = K_\mathrm{mol}C_\mathrm{mol} = K_\mathrm{mol}\frac{N_{\mathrm{free}}}{N_AV_\mathrm{cv}} = a\,N_{\mathrm{free}},
  \label{eq:KC_lumped_mol}
\end{equation}
leading to a relation between $K_\mathrm{mol}$ and $a$ as
\begin{equation}
    K_\mathrm{mol} = aN_AV_\mathrm{cv}.
  \label{eq:KC_lumped_mol}
\end{equation}
As a result, $K_\mathrm{mol}\sim2.4\times10^5\,$M$^{-1}$ was obtained for Fig.\,\ref{fig:md_sim}. Moreover, MD simulations with more water molecules (6944 water molecules) in a larger periodic simulation cell were also performed; see \textcolor{blue}{Supplementary Fig.\,8}. The closed-system Langmuir isotherm fitting yielded $N_\mathrm{site} = 40.07\,\pm\,5.33$ and $a = 0.02\,\pm\,0.007$. Further, the fitting $N_\mathrm{ads,max}\sim28$, $N_\mathrm{tot,max}\sim145$, $C_\mathrm{mol,max}\sim12\,\upmu$M, and $V_\mathrm{cv}\sim2.0\times10^7\,\mathrm{nm^3}$. Hence, $K_\mathrm{mol}\sim2.5\times10^5\,\mathrm{M^{-1}}$ was obtained and is almost the same as the value from the simulation results using a smaller simulation domain, confirming that water molecules have little effect. 

\smallskip
\noindent\textbf{Chiral transfer matrix method (cTMM)} -- A cTMM solver for general reciprocal, chiral, and bianisotropic materials and their layered structures was developed to simulate and analyze sensor response~\cite{LindellEtAl1994,FanEtAl2025C}. Specifically, because of coupled electric and magnetic fields, the constitutive relations become $\mathbf{D} = \varepsilon\mathbf{E} + \xi\mathbf{H}$ and $\mathbf{B} = \zeta\mathbf{E} + \mu\mathbf{H}$, where $\mathbf{D}$ is displacement field, $\mathbf{E}$ is electric field, $\mathbf{H}$ is magnetic field, $\mathbf{B}$ is magnetic flux density, $\varepsilon$ is electric permittivity, $\mu$ is magnetic permeability, and $\xi$, $\zeta$ are two magnetoelectric coefficients describing the coupling between electric and magnetic components. Alternatively,  $\xi$ and $\zeta$ can be expressed as $\sqrt{\varepsilon_0\mu_0}(\chi - j\kappa)$ and  $\sqrt{\varepsilon_0\mu_0}(\chi + j\kappa)$, respectively, where $\varepsilon_0$ is vacuum permittivity, $\mu_0$ is vacuum permeability, and $\chi$ and $\kappa$ are two unitless magnetoelectric coefficients. For reciprocal and chiral materials, $\chi$ is zero and $\kappa$ is called the chirality parameter. 

Further, \textbf{E}(\textbf{r}) and \textbf{H}(\textbf{r}) fields of the incident wave with an angular frequency $\omega$ propagating along $z$ axis can be expressed as
\begin{align}
    \mathbf{E}(\mathbf{r}) &= \mathbf{e}(z)e^{jq(x\mathrm{cos}\psi + y\mathrm{sin}\psi)},\\
    \mathbf{H}(\mathbf{r}) &= \mathbf{h}(z)e^{jq(x\mathrm{cos}\psi + y\mathrm{sin}\psi)},
\end{align}
where $q = k_0n_1\mathrm{sin}\theta_\mathrm{i}$ is real-valued wavenumber with $k_0 = \omega\sqrt{\varepsilon_0\mu_0}$ , $\psi$ is azimuthal angle, $\theta_\mathrm{i}$ is incident angle, $n_1$ is refractive index of incident light medium, $\textbf{e}(z)$ is the unit vector expressed as $\mathbf{e_x}(z)\hat{\mathbf{x}} + \mathbf{e_y}(z)\hat{\mathbf{y}} + \mathbf{e_z}(z)\hat{\mathbf{z}}$, and $\mathbf{h}(z)$ is the unit vector as $\mathbf{h_x}(z)\hat{\mathbf{x}} + \mathbf{h_y}(z)\hat{\mathbf{y}} + \mathbf{h_z}(z)\hat{\mathbf{z}}$. For $z$-components, coupled electric and magnetic fields in constitutive relations lead to $\mathbf{e_z}(z) = \nu_{zx}^{ee}\mathbf{e_x}(z) + \nu_{zy}^{ee}\mathbf{e_y}(z) + \nu_{zx}^{eh}\mathbf{h_x}(z) + \nu_{zy}^{eh}\mathbf{h_y}(z)$ and $\mathbf{h_z}(z) = \nu_{zx}^{he}\mathbf{e_x}(z) + \nu_{zy}^{he}\mathbf{e_y}(z) + \nu_{zx}^{hh}\mathbf{h_x}(z) + \nu_{zy}^{hh}\mathbf{h_y}(z)$. 
Solving Maxwell's equations and matching boundary conditions yields
\begin{align}
    \nu_{zx}^{ee} = -\frac{\mu_{zz}\varepsilon_{zx} - \xi_{zz}[\zeta_{zx} + (q/\omega)\mathrm{sin}\psi]}{\varepsilon_{zz}\mu_{zz} - \xi_{zz}\zeta_{zz}},
    \nu_{zy}^{ee} = -\frac{\mu_{zz}\varepsilon_{zy} - \xi_{zz}[\zeta_{zy} - (q/\omega)\mathrm{cos}\psi]}{\varepsilon_{zz}\mu_{zz} - \xi_{zz}\zeta_{zz}}, \\
    \nu_{zx}^{eh} = \frac{\xi_{zz}\mu_{zx} - \mu_{zz}[\xi_{zx} - (q/\omega)\mathrm{sin}\psi]}{\varepsilon_{zz}\mu_{zz} - \xi_{zz}\zeta_{zz}}, 
    \nu_{zy}^{eh} = \frac{\xi_{zz}\mu_{zy} - \mu_{zz}[\xi_{zy} + (q/\omega)\mathrm{cos}\psi]}{\varepsilon_{zz}\mu_{zz} - \xi_{zz}\zeta_{zz}}, \\
    \nu_{zx}^{he} = \frac{\zeta_{zz}\varepsilon_{zx} - \varepsilon_{zz}[\zeta_{zx} + (q/\omega)\mathrm{sin}\psi]}{\varepsilon_{zz}\mu_{zz} - \xi_{zz}\zeta_{zz}}, 
    \nu_{zy}^{he} = \frac{\zeta_{zz}\varepsilon_{zy} - \varepsilon_{zz}[\zeta_{zy} - (q/\omega)\mathrm{cos}\psi]}{\varepsilon_{zz}\mu_{zz} - \xi_{zz}\zeta_{zz}}, \\
    \nu_{zx}^{hh} = -\frac{\varepsilon_{zz}\mu_{zx} - \zeta_{zz}[\xi_{zx} - (q/\omega)\mathrm{sin}\psi]}{\varepsilon_{zz}\mu_{zz} - \xi_{zz}\zeta_{zz}}, 
    \nu_{zy}^{hh} = -\frac{\varepsilon_{zz}\mu_{zy} - \zeta_{zz}[\xi_{zy} + (q/\omega)\mathrm{cos}\psi]}{\varepsilon_{zz}\mu_{zz} - \xi_{zz}\zeta_{zz}},
\end{align}
where $\varepsilon_{ij}, \mu_{ij}, \xi_{ij}, \zeta_{ij}, i,j \in \{x,y,z\}$ are matrix elements of $\varepsilon, \mu, \xi$, and $\zeta$, respectively, for materials in each layer. Then, matrices for $m$-th layer with a thickness $d_m$, $P_{m}$ and $M_m$, can be expressed as
\begin{equation}
\begin{aligned}
&P_{m} = \omega
    \left (
    \begin{bmatrix}
        \zeta_{yx} & \zeta_{yy} & \mu_{yx} & \mu_{yy} \\
        -\zeta_{xx} & -\zeta_{xy} & -\mu_{xx} & -\mu_{xy} \\
        -\varepsilon_{yx} & -\varepsilon_{yy} & -\xi_{yx} & -\xi_{yy} \\
        \varepsilon_{xx} & \varepsilon_{xy} & \xi_{xx} & \xi_{xy} 
    \end{bmatrix} + 
    \right.
    \\&
    \left.
    \begin{bmatrix}
        \zeta_{yz} + \frac{q}{\omega}\mathrm{cos}\psi & 0 & 0 & 0 \\
        0 & -\zeta_{xz} + \frac{q}{\omega}\mathrm{sin}\psi & 0 & 0 \\
        0 & 0 & -\varepsilon_{yz} & 0 \\
        0 & 0 & 0 & \varepsilon_{xz} 
    \end{bmatrix}
    \begin{bmatrix}
        1 & 1 & 1 & 1 \\
        1 & 1 & 1 & 1 \\
        1 & 1 & 1 & 1 \\
        1 & 1 & 1 & 1  
    \end{bmatrix}
    \begin{bmatrix}
        \nu_{zx}^{ee} & 0 & 0 & 0 \\
        0 & \nu_{zy}^{ee} & 0 & 0 \\
        0 & 0 & \nu_{zx}^{eh} & 0 \\
        0 & 0 & 0 & \nu_{zy}^{eh} 
    \end{bmatrix} + 
    \right.
    \\&
    \left.
    \begin{bmatrix}
        \mu_{yz} & 0 & 0 & 0 \\
        0 & -\mu_{xz} & 0 & 0 \\
        0 & 0 & -\xi_{yz} + \frac{q}{\omega}\mathrm{cos}\psi & 0 \\
        0 & 0 & 0 & \xi_{xz} + \frac{q}{\omega}\mathrm{sin}\psi
    \end{bmatrix}
    \begin{bmatrix}
        1 & 1 & 1 & 1 \\
        1 & 1 & 1 & 1 \\
        1 & 1 & 1 & 1 \\
        1 & 1 & 1 & 1  
    \end{bmatrix}
    \begin{bmatrix}
        \nu_{zx}^{he} & 0 & 0 & 0 \\
        0 & \nu_{zy}^{he} & 0 & 0 \\
        0 & 0 & \nu_{zx}^{hh} & 0 \\
        0 & 0 & 0 & \nu_{zy}^{hh} 
    \end{bmatrix} + 
    \right )
\end{aligned}
\end{equation}
and
\begin{align}
    M_m = e^{j P_m d_m}.
\end{align}
In addition, matrices for the incident and transmission layers in terms of $s$- and $p$-polarized waves with the incident angle $\theta_\mathrm{i}$ and output angle $\theta_\mathrm{tr}$, $K_\mathrm{inc}$ and $K_\mathrm{tr}$, can be expressed as
\begin{align*}
    K_\mathrm{inc} = 
    \begin{bmatrix}
        -\mathrm{sin} \psi & -\mathrm{cos} \psi \mathrm{cos} \theta_\mathrm{i} & -\mathrm{sin} \psi & \mathrm{cos} \psi \mathrm{cos} \theta_\mathrm{i} \\
        \mathrm{cos} \psi & -\mathrm{sin} \psi \mathrm{cos} \theta_\mathrm{i} & \mathrm{cos} \psi & \mathrm{sin} \psi \mathrm{cos} \theta_\mathrm{i} \\
        -\left(\frac{n_1}{\eta_0}\right)\mathrm{cos}\psi \mathrm{cos} \theta_\mathrm{i} & \left(\frac{n_1}{\eta_0}\right)\mathrm{sin}\psi & \left(\frac{n_1}{\eta_0}\right)\mathrm{cos}\psi \mathrm{cos} \theta_\mathrm{i} & \left(\frac{n_1}{\eta_0}\right)\mathrm{sin}\psi \\
        -\left(\frac{n_1}{\eta_0}\right)\mathrm{sin}\psi \mathrm{cos} \theta_\mathrm{i} & -\left(\frac{n_1}{\eta_0}\right)\mathrm{cos}\psi & \left(\frac{n_1}{\eta_0}\right)\mathrm{sin}\psi \mathrm{cos} \theta_\mathrm{i} & -\left(\frac{n_1}{\eta_0}\right)\mathrm{cos}\psi 
    \end{bmatrix}
\end{align*}
and
\begin{align*}
    K_\mathrm{tr} = 
    \begin{bmatrix}
        -\mathrm{sin} \psi & -\mathrm{cos} \psi \mathrm{cos} \theta_\mathrm{tr} & -\mathrm{sin} \psi & \mathrm{cos} \psi \mathrm{cos} \theta_\mathrm{tr} \\
        \mathrm{cos} \psi & -\mathrm{sin} \psi \mathrm{cos} \theta_\mathrm{tr} & \mathrm{cos} \psi & \mathrm{sin} \psi \mathrm{cos} \theta_\mathrm{tr} \\
        -\left(\frac{n_2}{\eta_0}\right)\mathrm{cos}\psi \mathrm{cos} \theta_\mathrm{tr} & \left(\frac{n_2}{\eta_0}\right)\mathrm{sin}\psi & \left(\frac{n_2}{\eta_0}\right)\mathrm{cos}\psi \mathrm{cos} \theta_\mathrm{tr} & \left(\frac{n_2}{\eta_0}\right)\mathrm{sin}\psi \\
        -\left(\frac{n_2}{\eta_0}\right)\mathrm{sin}\psi \mathrm{cos} \theta_\mathrm{tr} & -\left(\frac{n_2}{\eta_0}\right)\mathrm{cos}\psi & \left(\frac{n_2}{\eta_0}\right)\mathrm{sin}\psi \mathrm{cos} \theta_\mathrm{tr} & -\left(\frac{n_2}{\eta_0}\right)\mathrm{cos}\psi 
    \end{bmatrix}.
\end{align*}
Hence, the overall transfer matrix $Q = K_\mathrm{tr}^{-1}\bigg(\displaystyle\prod_{m=1}^{N}M_m\bigg)K_\mathrm{inc}$, where $N$ is the total number of layers. The input amplitudes for s- and p-polarized waves, $a_\mathrm{s}$ and $a_\mathrm{p}$, the reflection amplitudes for s- and p-polarized waves, $r_\mathrm{s}$ and $r_\mathrm{p}$, and the transmission amplitudes for s- and p-polarized waves, $t_\mathrm{s}$ and $t_\mathrm{r}$, are related through $Q$ as
\begin{align}
    \begin{bmatrix}
        t_\mathrm{s} \\
        t_\mathrm{p} \\
        0 \\
        0 \\
    \end{bmatrix}  = Q
    \begin{bmatrix}
        a_\mathrm{s} \\
        a_\mathrm{p} \\
        r_\mathrm{s} \\
        r_\mathrm{p} \\
    \end{bmatrix} = 
    \begin{bmatrix}
        Q_{00} & Q_{01} & Q_{02} & Q_{03} \\
        Q_{10} & Q_{11} & Q_{12} & Q_{13} \\
        Q_{20} & Q_{21} & Q_{22} & Q_{23} \\
        Q_{30} & Q_{31} & Q_{32} & Q_{33} 
    \end{bmatrix} 
    \begin{bmatrix}
        a_\mathrm{s} \\
        a_\mathrm{p} \\
        r_\mathrm{s} \\
        r_\mathrm{p} \\
    \end{bmatrix}.
\end{align}
The reflection and transmission coefficients under different combinations of $s$ and $p$ linear polarizations, $r_\mathrm{ss}, r_\mathrm{sp}, r_\mathrm{ps}, r_\mathrm{pp}, t_\mathrm{ss}, t_\mathrm{sp}, t_\mathrm{ps},$ and $t_\mathrm{pp}$, are related through 
\begin{align}
    \begin{bmatrix}
        r_\mathrm{s} \\
        r_\mathrm{p} \\
    \end{bmatrix}  =
    \begin{bmatrix}
        r_\mathrm{ss} & r_\mathrm{sp} \\
        r_\mathrm{ps} & r_\mathrm{pp}
    \end{bmatrix}
    \begin{bmatrix}
        a_\mathrm{s} \\
        a_\mathrm{p} \\
    \end{bmatrix},
\end{align}
and 
\begin{align}
    \begin{bmatrix}
        t_\mathrm{s} \\
        t_\mathrm{p} \\
    \end{bmatrix}  =
    \begin{bmatrix}
        t_\mathrm{ss} & t_\mathrm{sp} \\
        t_\mathrm{ps} & t_\mathrm{pp}
    \end{bmatrix} 
    \begin{bmatrix}
        a_\mathrm{s} \\
        a_\mathrm{p} \\
    \end{bmatrix},
\end{align}
and their values can be calculated based on the matrix elements of $Q$ using
\begin{align}
    r_\mathrm{ss} = \frac{Q_{30}Q_{23} - Q_{33}Q_{20}}{Q_{22}Q_{23} - Q_{23}Q_{32}}, \quad
    r_\mathrm{sp} = \frac{Q_{31}Q_{23} - Q_{33}Q_{21}}{Q_{22}Q_{23} - Q_{23}Q_{32}}, \\
    r_\mathrm{ps} = \frac{Q_{32}Q_{20} - Q_{30}Q_{22}}{Q_{22}Q_{23} - Q_{23}Q_{32}}, \quad
    r_\mathrm{pp} = \frac{Q_{32}Q_{21} - Q_{31}Q_{22}}{Q_{22}Q_{23} - Q_{23}Q_{32}}, \\
    t_\mathrm{ss} = Q_{00} + Q_{02}r_\mathrm{ss} + Q_{03}r_\mathrm{ps}, \quad
    t_\mathrm{sp} = Q_{01} + Q_{02}r_\mathrm{sp} + Q_{03}r_\mathrm{pp}, \\
    t_\mathrm{ps} = Q_{10} + Q_{12}r_\mathrm{ss} + Q_{13}r_\mathrm{ps}, \quad
    t_\mathrm{pp} = Q_{11} + Q_{12}r_\mathrm{sp} + Q_{13}r_\mathrm{pp}. 
\end{align}
Further, the coefficients under L and R circular polarizations can be obtained from
\begin{align}
    r_\mathrm{LL} = -\frac{r_\mathrm{ss} + r_\mathrm{pp} + j(r_\mathrm{sp} - r_\mathrm{ps})}{2}, \quad
    r_\mathrm{LR} = \frac{r_\mathrm{ss} - r_\mathrm{pp} - j(r_\mathrm{sp} + r_\mathrm{ps})}{2}, \\
    r_\mathrm{RL} = \frac{r_\mathrm{ss} - r_\mathrm{pp} + j(r_\mathrm{sp} + r_\mathrm{ps})}{2}, \quad
    r_\mathrm{RR} = -\frac{r_\mathrm{ss} + r_\mathrm{pp} - j(r_\mathrm{sp} - r_\mathrm{ps})}{2}, \\
    t_\mathrm{LL} = \frac{t_\mathrm{ss} + t_\mathrm{pp} + j(t_\mathrm{sp} - t_\mathrm{ps})}{2}, \quad
    t_\mathrm{LR} = -\frac{t_\mathrm{ss} - t_\mathrm{pp} - j(t_\mathrm{sp} + t_\mathrm{ps})}{2}, \\
    t_\mathrm{RL} = -\frac{t_\mathrm{ss} - t_\mathrm{pp} + j(t_\mathrm{sp} + t_\mathrm{ps})}{2}, \quad
    t_\mathrm{RR} = \frac{t_\mathrm{ss} + t_\mathrm{pp} - j(t_\mathrm{sp} - t_\mathrm{ps})}{2}.
\end{align}
Hence, the power reflectance $R$, transmittance $T$, and attenuation $A$ for any circular polarization combination can be calculated using $R = |r|^2$, $T = \frac{n_2\mathrm{Re\{cos}\theta_\mathrm{tr}\}}{n_1\mathrm{cos}\theta_\mathrm{tr}}|t|^2$, and $A=-\mathrm{log}_{10}T$. CD is thus $A_\mathrm{L} - A_\mathrm{R}$ and average absorption is $(A_\mathrm{L} + A_\mathrm{R})/2$, where $A_\mathrm{L} = -\mathrm{log}_{10}(T_\mathrm{LL} + T_\mathrm{RL})$ is the attenuation for a left-handed circularly polarized incident light and $A_\mathrm{R} = -\mathrm{log}_{10}(T_\mathrm{RR} + T_\mathrm{LR})$ is the attenuation for a right-handed circularly polarized incident light. 

\smallskip
\noindent\textbf{Optical modeling of aligned CNTs with adsorbed chiral molecules} -- The relative electric permittivity of an aligned CNT film, $\varepsilon_\mathrm{CNT}$, can be expressed as
\begin{align}
    \varepsilon_\mathrm{CNT} = 
    \begin{bmatrix}
        \varepsilon_\mathrm{\parallel} & 0 & 0 \\
        0 & \varepsilon_\mathrm{\perp} & 0 \\
        0 & 0 & \varepsilon_\mathrm{\perp}
    \end{bmatrix},
\end{align}
where $\varepsilon_\mathrm{\parallel}$ and $\varepsilon_\mathrm{\perp}$ are the dielectric functions along and perpendicular to the alignment direction and their spectral response in a wavelength range from 200 to 800\,nm were modeled as a sum of Voigt functions~\cite{FanEtAl2025NC}. \textcolor{blue}{Supplementary Fig.\,9a} displays the real and imaginary parts of $\varepsilon_\mathrm{\parallel}$ and $\varepsilon_\mathrm{\perp}$ by simultaneously fitting CD and absorption spectra, which are shown in \textcolor{blue}{Supplementary Fig.\,9b} and \textcolor{blue}{Supplementary Fig.\,9c}, respectively. When there is an in-plane rotation angle, $\theta_\textrm{rot}$, between the CNT alignment direction in one layer and the CNT alignment direction in a reference layer, the rotated dielectric function, $\varepsilon_\mathrm{CNT, rot}$, becomes
\begin{align}
    \varepsilon_\mathrm{CNT, rot} = 
    \begin{bmatrix}
        \textrm{cos}\theta_\textrm{rot} & -\textrm{sin}\theta_\textrm{rot} & 0  \\
        \textrm{sin}\theta_\textrm{rot} & \textrm{cos}\theta_\textrm{rot} & 0 \\
        0 & 0 & 1 \\
    \end{bmatrix}^{-1}
    \begin{bmatrix}
        \varepsilon_{\parallel} & 0 & 0  \\
        0 & \varepsilon_{\perp} & 0 \\
        0 & 0 & \varepsilon_{\perp} \\
    \end{bmatrix}
    \begin{bmatrix}
        \textrm{cos}\theta_\textrm{rot} & -\textrm{sin}\theta_\textrm{rot} & 0  \\
        \textrm{sin}\theta_\textrm{rot} & \textrm{cos}\theta_\textrm{rot} & 0 \\
        0 & 0 & 1 \\
    \end{bmatrix}.
\end{align}

The CNTs inside aligned films contained a racemic mixture with a zero chirality parameter. The adsorption of chiral molecules on aligned CNTs induced a nonzero $\kappa_\mathrm{eff}$ only along the CNT alignment direction, denoted as diag($\kappa_\mathrm{eff}$, 0, 0). $\kappa_\mathrm{eff} (\omega)$ as a function of angular frequency $\omega$ was modeled using a standard Condon model as
\begin{align}
    \kappa_\mathrm{eff} = \frac{\kappa_\mathrm{amp}\omega R}{\omega_0^2 - \omega^2 - j\omega\gamma},
\end{align}
where $\kappa_\mathrm{amp}$ is a scaling amplitude factor, $R$ is the rotational strength, $\omega_0$ is the resonance frequency, and $\gamma$ is the damping rate. When the alignment rotated, the diagonal chirality parameter matrix was transformed similarly to $\varepsilon_\mathrm{CNT, rot}$. Further, since $\kappa_\mathrm{eff}$ originated from the adsorption coverage of chiral molecules, $\kappa_\mathrm{amp}$ was modeled as a function of molar concentration $C_\mathrm{mol}$ as
\begin{align}
    \kappa_\mathrm{eff} = h_\kappa F_\mathrm{s}\Delta\mathrm{CD}_\mathrm{trans}\frac{K_\mathrm{mol}C_\mathrm{mol}}{1+K_\mathrm{mol}C_\mathrm{mol}}.
\end{align}
Here, $F_\mathrm{s}$ is a universal scaling coefficient and was assumed to be the same for left-handed and right-handed chiral molecules. $h_\kappa$ is the sign operator with $h_\kappa = 1$ for the molecular transition with a negative CD and $h_\kappa = -1$ for the molecular transition with a positive CD under the convention used in the cTMM solver. Although glucose enantiomers have featureless CD spectra around the ultraviolet peak at 267\,nm wavelength, the CD spectra in the shorter wavelength show a positive peak for D-glucose and a negative peak for L-glucose~\cite{MatsuoEtAl2004CR}. Similarly, the CD spectral peak is positive for D-alanine and is negative for L-alanine~\cite{MeinertEtAl2022NC}. $\Delta\mathrm{CD}_\mathrm{trans}, K_\mathrm{mol}, C_\mathrm{mol}$ were all obtained by fitting experimental results as shown in Fig.\,\ref{fig:exp_results}. Following these conventions and expressions, all simulation sensing curves show excellent agreement with experimental results. 

\section*{Data availability}
The data that support the findings of this study are available from the corresponding author upon request. 

\section*{Acknowledgments}
H.X., J.F., and W.G. acknowledge support from the National Science Foundation through Grants No.\,2321366 and 2230727. Y.W. acknowledges the support provided by the University of Utah Research Foundation (UURF) through the Ascender Grant. C.G. is supported by the Jack Kilby/Texas Instruments Endowed Faculty Fellowship. J.L. acknowledges the financial support from the Air Force Office of Scientific Research, Multidisciplinary University Research Initiatives (MURI) Program under award number FA9550-23-1-0311 for the conceptual design of molecular dynamics simulation. S.R. acknowledges the financial support from the National Science Foundation, under the award number CBET-2312304, for the molecular dynamics simulation methodology development. 

\section*{Author Contributions Statement}
W.G. conceived the idea and led the project. H.X. performed the experiments with the help of J.F., and theoretical modeling and analysis with the help of C.G, under the supervision of W.G. Z.A.R.B. and S.R. performed the MD simulation under the supervision of J.L. M.M. performed the Raman spectroscopy under the supervision of Y.W. 

\section*{Competing Interests Statement}
The authors declare no competing interests.


\end{document}


\maketitle

\newpage
\begin{figure}[hbt]
    \centering
    \includegraphics[width = 0.40\textwidth]{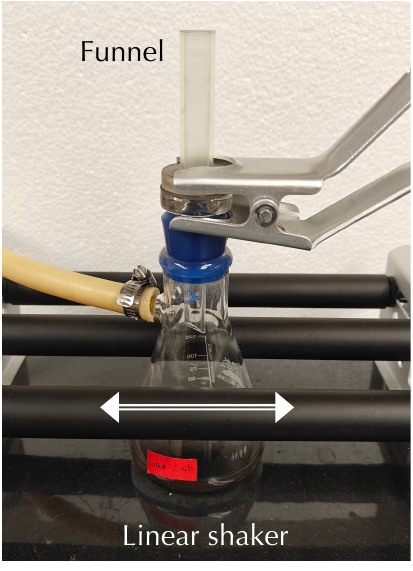}
    \caption{A photo of the shaking-assisted vacuum filtration setup.}
    \label{SI_fig:savf}
\end{figure}

\newpage
\begin{figure}[hbt]
    \centering
    \includegraphics[width = 0.6\textwidth]{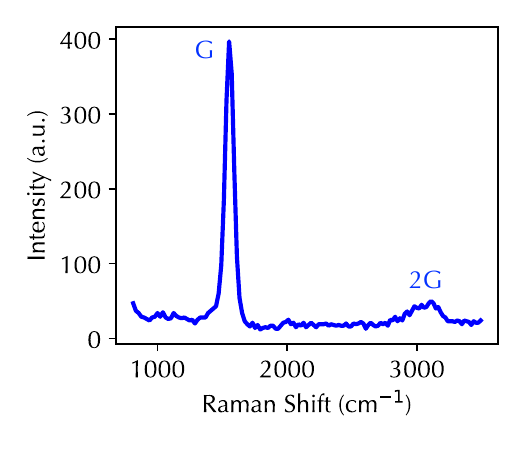}
    \caption{Raman spectrum of a prepared aligned CNT film.}
    \label{SI_fig:raman}
\end{figure}

\newpage
\begin{figure}[hbt]
    \centering
    \includegraphics[width = 0.3\textwidth]{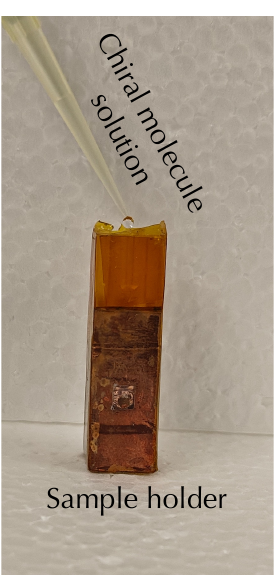}
    \caption{A photo of the sensor characterization apparatus.}
    \label{SI_fig:sensing_setup}
\end{figure}

\newpage
\begin{figure}[hbt]
    \centering
    \includegraphics[width = 0.9\textwidth]{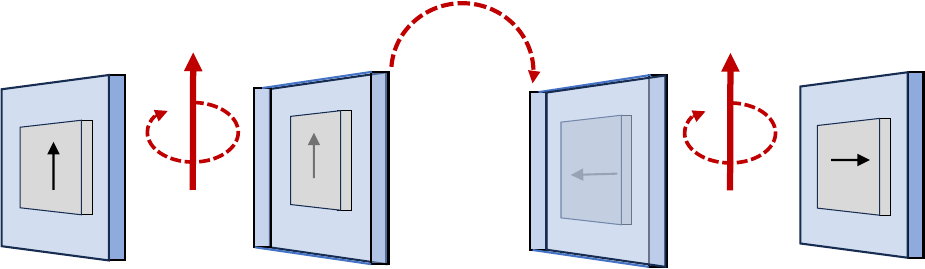}
    \caption{Schematic of four-configuration measurements of CD spectra.}
    \label{SI_fig:sensing_setup}
\end{figure}

\newpage
\begin{figure}[hbt]
    \centering
    \includegraphics[width = 0.6\textwidth]{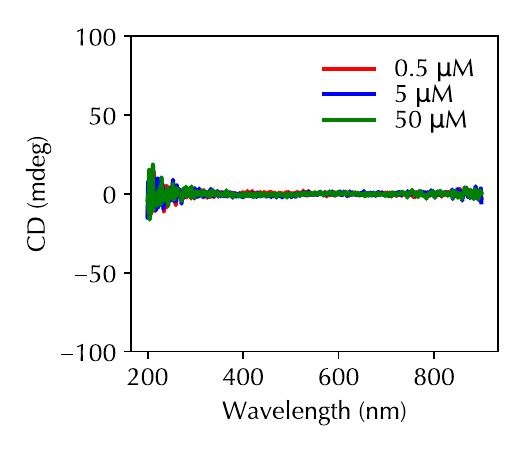}
    \caption{CD spectra of pure glucose solutions at 0.5\,$\upmu$M (red line), 5\,$\upmu$M (blue line), and 50\,$\upmu$M (green line).}
    \label{SI_fig:bare_glucose}
\end{figure}

\newpage
\begin{figure}[hbt]
    \centering
    \includegraphics[width = \textwidth]{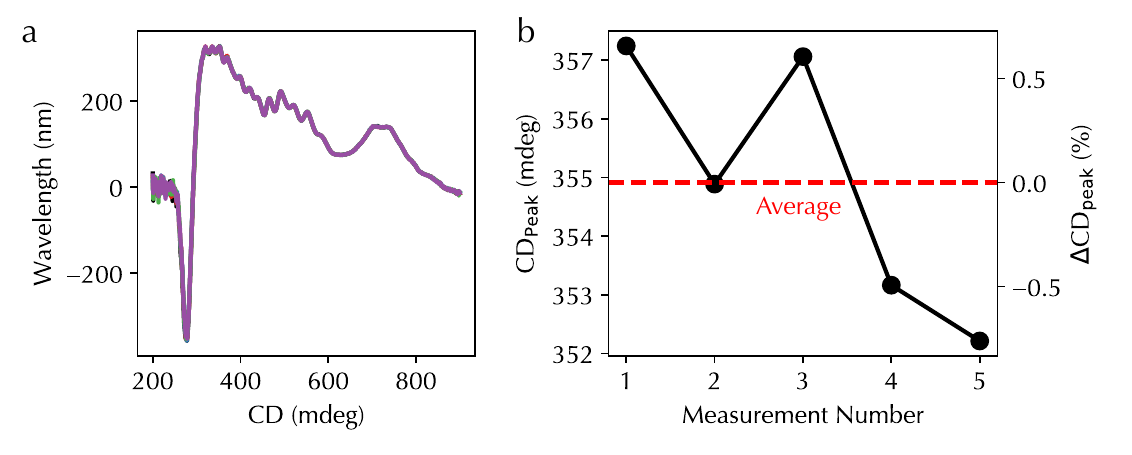}
    \caption{(a)~Repeated CD spectra measurements of the same twisted aligned CNTs. All lines overlap. (b)~CD$_\mathrm{peak}$ and $\Delta\mathrm{CD}_\mathrm{peak}$ of the repeated measurements (black dots and line). The red dashed line indicates the average CD$_\mathrm{peak}$ and $\Delta\mathrm{CD}_\mathrm{peak}=0$. }
    \label{SI_fig:exp_repeat}
\end{figure}

\newpage
\begin{figure}[hbt]
    \centering
    \includegraphics[width = \textwidth]{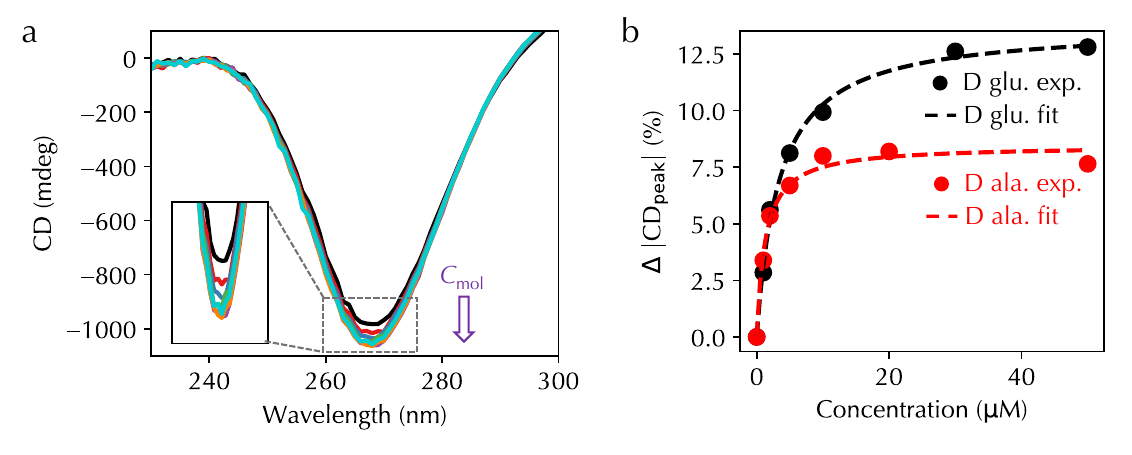}
    \caption{(a)~Experimental CD spectra of an L-twist sensor in (a)~the L-alanine solution with the concentration from $0\,\upmu$M to $50\,\upmu$M (from black to cyan lines). (b)~Comparison of the experimental (dots) and fitting (dashed lines) sensing curves for the L-twist sensor in the L-glucose (black colors) and L-alanine (red colors) solutions. }
    \label{SI_fig:alanine_compare}
\end{figure}

\newpage
\begin{figure}[hbt]
    \centering
    \includegraphics[width = 0.6\textwidth]{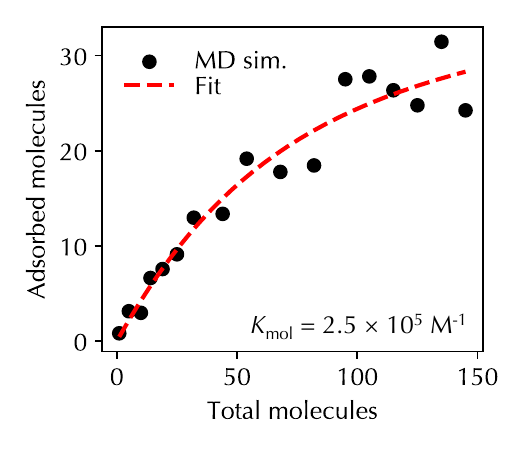}
    \caption{MD simulation (black dots) and fitting (red dashed line) number of adsorbed glucose molecules with respect to the total number of glucose molecules in the system under a total number of 6944 water molecules.}
    \label{SI_fig:md_sim}
\end{figure}

\newpage
\begin{figure}[hbt]
    \centering
    \includegraphics[width = \textwidth]{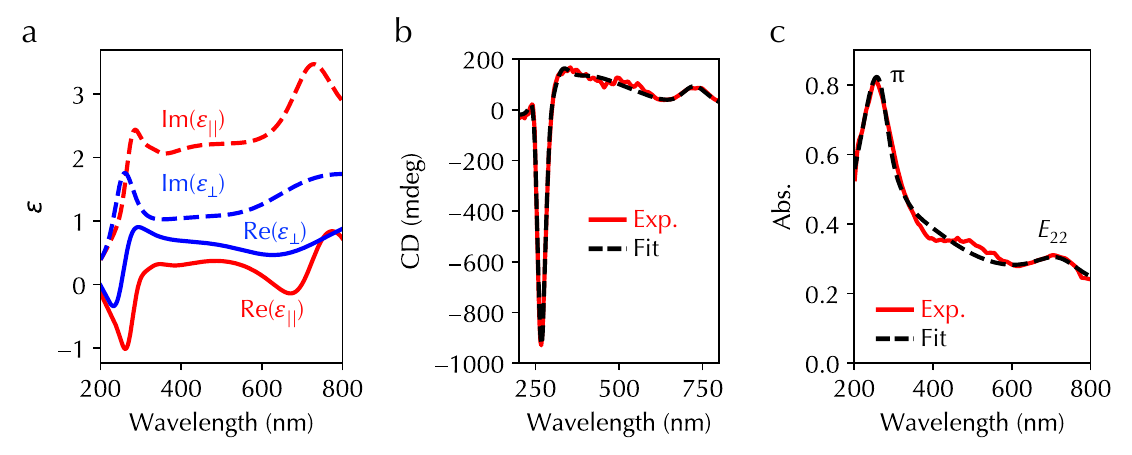}
    \caption{(a)~Real part of the dielectric function along the CNT axis (red solid line), imaginary part of the dielectric function along the CNT axis (red dashed line), real part of the dielectric function perpendicular to the CNT axis (blue solid line), and imaginary part of the dielectric function perpendicular to the CNT axis (blue dashed line). (b)~Experimental (red solid line) and fitting (black dashed line) CD spectra of a two-layer twisted aligned CNT sample. (c)~The corresponding experimental (red solid line) and fitting (black dashed line) absorption spectra. }
    \label{SI_fig:cnt_dielectric}
\end{figure}

\newpage
\begin{figure}[hbt]
    \centering
    \includegraphics[width = \textwidth]{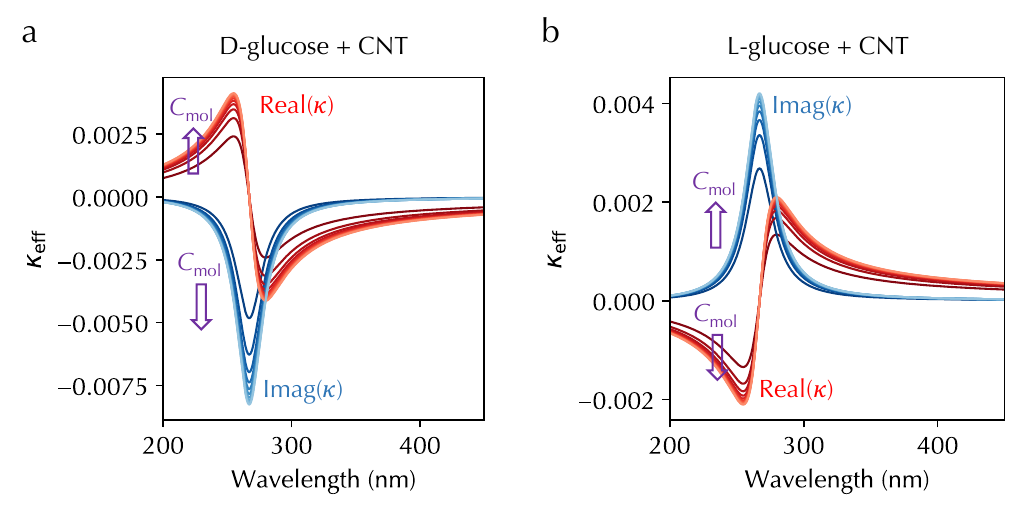}
    \caption{(a)~Real parts (dark to light red lines) and (b)~imaginary parts (dark to light blue lines) of the fitting $\kappa_\mathrm{eff}$ along the CNT axis at various concentrations.}
    \label{SI_fig:fit_kappa}
\end{figure}

\newpage
\begin{table}[hbt]
    \centering
    \caption{\bf \centering Glucose enantioselective sensing in different platforms using CD spectroscopy}
    \begin{tabular}{|c|c|c|c|}
    \hline
    
    \makecell{Sensor platform}  & \makecell{LOD} & \makecell{Molecule delivery} & Reference \\
    \hline

    \makecell{Moir\'e chiral metamaterials}  & \makecell{10\,mM (no bubble)} & Microfluidic cell & Ref.\,\cite{LiuEtAl2021N}\\
    \hline

    \makecell{Chiral metamaterials}  & \makecell{0.03\,mM} & Dried sample & Ref.\,\cite{XuEtAl2024AFM}\\
    \hline

    \makecell{Chiral gold nanoparticles}  & \makecell{10s$-$100s of mM (estimated)} & in situ & Ref.\,\cite{KimEtAl2022N}\\
    \hline

    \makecell{Nonlinear chiral metamaterials}  & \makecell{$>10$\,mM (estimated)} & N.A. & Ref.\,\cite{LeeEtAl2022AFM}\\
    \hline

    \makecell{Biomimetic receptor}  & \makecell{sub-mM} & Solution titration & Ref.\,\cite{TromansEtAl2020CS}\\
    \hline

    \makecell{Our twisted aligned CNTs} & $ \sim0.3\,\upmu$M & in situ &This work\\
    \hline

    \end{tabular}
      \label{tab:adsorption_const}
\end{table}

\newpage
\begin{table}[hbt]
    \centering
    \caption{\bf \centering  Glucose adsorption in different platforms}
    \begin{tabular}{|c|c|c|c|}
    \hline
    
    Adsorbent platform   & Equilibrium constant ($\mathrm{M}^{-1}$)  & Half-saturation concentration ($1/K$) & Reference \\
    \hline

    \makecell{Glucose oxidase}   & $\sim110$ & $9.1\,$mM  & Ref.\,\cite{YangEtAl2016SAC}\\
    \hline

    \makecell{Alkali-activated \\ microporous carbon}   & $\sim220$ & $4.5\,$mM  & Ref.\,\cite{YabushitaEtAl2014C}\\
    \hline

    \makecell{Diboronic acid}   & $\sim2000 - \sim6500$ & $\sim0.15 - \sim0.5\,$mM  & Ref.\,\cite{NanEtAl2023B}\\
    \hline

    \makecell{Glucose oxidase \\ functionalized CNTs}   & $\sim830$ & $\sim1.2\,$mM  & Ref.\,\cite{BaroneEtAl2005NM}\\
    \hline

    \makecell{Concanavalin A}   & $\sim500$ & $2.0\,$mM  & Ref.\,\cite{TooneEtAl1994COSB}\\
    \hline

    \makecell{Synthetic lectins}   & $\sim250$ & $4.0\,$mM  & Ref.\,\cite{RiosEtAl2017CS}\\
    \hline
    
    \makecell{Biomimetic receptor}   & $\sim1.8\times10^4$ & $55.6\,\upmu$M  & Ref.\,\cite{TromansEtAl2019NC}\\
    \hline

    \makecell{Glucose-binding protein}   & $\sim7700$ & $130\,\upmu$M  & Ref.\,\cite{VeetilEtAl2010BB}\\
    \hline

    \makecell{Thermophilic glucose \\ binding protein}   & $\sim1.3\times10^7$ & $0.08\,\upmu$M  & Ref.\,\cite{CuneoEtAl2006JMB}\\
    \hline

    \makecell{Graphene quantum dots}   & $\sim4.0\times10^4$ & $25\,\upmu$M  & Ref.\,\cite{RajmaneEtAl2025SR}\\
    \hline

    \makecell{Our twisted aligned CNTs}   & $>10^5$ & $ <10\,\upmu$M  & This work\\
    \hline

    \end{tabular}
      \label{tab:adsorption_const}
\end{table}
